\documentclass[manuscript]{aastex}

\newcommand{\Lsun}{$L_{\odot}$}
\newcommand{\ccm}{cm$^3$}
\newcommand{\nhhh}{NH$_3$}
\newcommand{\kms}{km s$^{-1}$}

\shorttitle{A Sensitive Search for SiO Masers in SRF}
\shortauthors{Zapata et al.}
\def\folio{\ifnum\pageno=1\nopagenumbers\else\number\pageno\fi}

\def\lax    {\ifmmode{_<\atop^{\sim}}\else{${_<\atop^{\sim}}$}\fi}
\def\gax    {\ifmmode{_>\atop^{\sim}}\else{${_>\atop^{\sim}}$}\fi}
\newbox\grsign      \setbox\grsign=\hbox{$>$}
\newdimen\grdimen   \grdimen=\ht\grsign
\newbox\simgreatbox \setbox\simgreatbox=\hbox{\raise.5ex\hbox{$>$}\llap
                        {\lower.5ex\hbox{$\sim$}}}\ht1=\grdimen\dp1=0pt
\newbox\simlessbox  \setbox\simlessbox =\hbox{\raise.5ex\hbox{$<$}\llap
                        {\lower.5ex\hbox{$\sim$}}}\ht2=\grdimen\dp2=0pt

\def\d {\phantom{$0$}}
\def\dd {\phantom{$00$}}

\newbox\grsign \setbox\grsign=\hbox{$>$} \newdimen\grdimen \grdimen=\ht\grsign
\newbox\laxbox \newbox\gaxbox
\setbox\gaxbox=\hbox{\raise.5ex\hbox{$>$}\llap
     {\lower.5ex\hbox{$\sim$}}}\ht1=\grdimen\dp1=0pt
\setbox\laxbox=\hbox{\raise.5ex\hbox{$<$}\llap
     {\lower.5ex\hbox{$\sim$}}}\ht2=\grdimen\dp2=0pt
\def\gax{\mathrel{\copy\gaxbox}}
\def\lax{\mathrel{\copy\laxbox}}
\def\boxit#1    {\vbox{\hrule\hbox{\vrule\kern3pt
                  \vbox{\kern3pt#1\kern3pt}\kern3pt\vrule}\hrule}}
\def\h      {\ifmmode{^{\rm h}}\else{$^{\rm h}$}\fi}
\def\m      {\ifmmode{^{\rm m}}\else{$^{\rm m}$}\fi}
\def\s      {\ifmmode{^{\rm s}}\else{$^{\rm s}$}\fi}
\def\am     {\ifmmode {\rlap.}$\,$'$\,$\! \else ${\rlap.}$\,$'$\,$\!$\fi}
\def\decam     {\ifmmode {\rlap.}$\,$'$\,$\! \else ${\rlap.}$\,$'$\,$\!$\fi}
\def\decas    {\ifmmode{{\rlap.}{''}}\else{${\rlap.}{''}$}\fi}
\def\mum     {\ifmmode{\mu{\rm m}}\else{$\mu{\rm m}$}\fi}
\def\decam {\rlap . {}'}          
\def\am {\rlap . {}'}          
\def\s      {\ifmmode{^{\rm s}}\else{$^{\rm s}$}\fi}
\def\deg      {\ifmmode{^{\circ}}\else{$^{\circ}$}\fi}
\def\as     {\ifmmode {\rlap.}$\,$''$\,$\! \else ${\rlap.}$\,$''$\,$\!$\fi}
\def\decsec  {\ifmmode {\rlap.}$\,$^{s}$\,$\! \else ${\rlap.}$\,$^{s}$\,$\!$\fi}\def\decs  {\ifmmode {\rlap.}$\,$^{s}$\,$\! \else ${\rlap.}$\,$^{s}$\,$\!$\fi}

\def\kms    {\ifmmode{{\rm km~s}^{-1}}\else{km~s$^{-1}$}\fi}

\def\ccm    {cm$^{-3}$}

\def\Lsun   {$L_{\odot}$}

\def\Msun   {$M_{\odot}$}
\def\Mspy   {\ifmmode {M_{\odot} {\rm yr}^{-1}} \else $M_{\odot}$~yr$^{-1}$\fi}
\def\Mdot   {\ifmmode {\dot M} \else $\dot M$\fi}
\def\mhd    {\ifmmode {n_{{\rm H}_2}} \else $n_{{\rm H}_2}$\fi}
\def\mhcd   {\ifmmode {N_{{\rm H}_2}} \else $N_{{\rm H}_2}$\fi}

\def\El      {\ifmmode{E_{\ell}}\else{$E_{\ell}$}\fi}
\def\beam    {\ifmmode{\theta_{\rm B}}\else{$\theta_{\rm B}$}\fi}
\def\Jyb   {\ifmmode {{\rm Jy~beam}^{-1}} \else{Jy~beam$^{-1}$}\fi}
\def\mjyb   {\ifmmode {{\rm mJy~beam}^{-1}} \else{mJy~beam$^{-1}$}\fi}
\def\mujyb   {\ifmmode {\mu{\rm Jy~beam}^{-1}} \else{$\mu$Jy~beam$^{-1}$}\fi}
\def\Trot   {\ifmmode{T_{\rm rot}}\else$T_{\rm rot}$\fi}

\def\Teff   {\ifmmode{T_{\rm eff}}\else$T_{\rm eff}$\fi}

\def\ITRS   {\ifmmode{\smallint {\rm T}_{R}^{*}dv}\else{$\smallint
{\rm T}_{R}^{*}dv$}\fi}
\def\ITRS   {\ifmmode{\smallint {\rm T}_{R}^{*}dv}\else{$\smallint
{\rm T}_{R}^{*}dv$}\fi}
\def\ITAS   {\ifmmode{\smallint {\rm T}_{A}^{*}dv}\else{$\smallint
{\rm T}_{A}^{*}dv$}\fi}

\def\nhhh       {NH$_3$}

\def\lefttitle#1  {\noindent \hangindent=18.0pt \hangafter=1 {#1} \par}
\def\vol#1  {{\bf {#1}{\rm,}\ }}
\font\tenssb=cmssbx10
\font\tenbf=cmbx10
\font\sevenbf=cmbx8
\font\fivebf=cmbx6
\def\unetdemi    {\smallskipamount=6pt plus2pt minus2pt
                  \medskipamount=12pt plus4pt minus4pt
                  \bigskipamount=24pt plus8pt minus8pt
                  \normalbaselineskip=16pt plus0pt minus0pt
                  \normallineskip=2pt
                  \normallineskiplimit=0pt
                  \jot=6pt
                  {\def\smallskip {\vskip\smallskipamount}}
                  {\def\medskip   {\vskip\medskipamount}}
                  {\def\bigskip   {\vskip\bigskipamount}}
                  {\setbox\strutbox=\hbox{\vrule
                    height17.0pt depth7.0pt width 0pt}}
                  \parskip 12.0pt
                  \normalbaselines}
\def\smallerspace {\smallskipamount=3pt plus0pt minus0pt
                  \medskipamount=6pt plus0pt minus0pt
                  \bigskipamount=10.5pt plus0pt minus0pt
                  \normalbaselineskip=10.5pt plus0pt minus0pt
                  \normallineskip=1pt
                  \normallineskiplimit=0pt
                  \jot=3pt
                  {\def\smallskip {\vskip\smallskipamount}}
                  {\def\medskip   {\vskip\medskipamount}}
                  {\def\bigskip   {\vskip\bigskipamount}}
                  {\setbox\strutbox=\hbox{\vrule
                    height8.5pt depth3.5pt width 0pt}}
                  \parskip 0pt
                  \normalbaselines}
\def\memospace    {\smallskipamount=4pt plus1pt minus1pt
                  \medskipamount=6pt plus2pt minus2pt
                  \bigskipamount=14pt plus6pt minus6pt
                  \normalbaselineskip=14pt plus0pt minus0pt
                  \normallineskip=1pt
                  \normallineskiplimit=0pt
                  \jot=4pt
                  {\def\smallskip {\vskip\smallskipamount}}
                  {\def\medskip   {\vskip\medskipamount}}
                  {\def\bigskip   {\vskip\bigskipamount}}
                  {\setbox\strutbox=\hbox{\vrule
                    height17.0pt depth7.0pt width 0pt}}
                  \parskip 2.0pt
                  \normalbaselines}
\def\memowidespace    {\smallskipamount=5pt plus1pt minus1pt
                  \medskipamount=7.5pt plus2pt minus2pt
                  \bigskipamount=17.5pt plus6pt minus6pt
                  \normalbaselineskip=17.0pt plus0pt minus0pt
                  \normallineskip=1.25pt
                  \normallineskiplimit=0pt
                  \jot=5pt
                  {\def\smallskip {\vskip\smallskipamount}}
                  {\def\medskip   {\vskip\medskipamount}}
                  {\def\bigskip   {\vskip\bigskipamount}}
                  {\setbox\strutbox=\hbox{\vrule
                    height21.25pt depth8.75pt width 0pt}}
                  \parskip 2.5pt
                  \normalbaselines}
\begin{document}
\title{An Extensive, Sensitive Search for SiO Masers in
High- and Intermediate-Mass Star-Forming Regions}

\author{Luis A. Zapata\altaffilmark{1}, Karl Menten\altaffilmark{1},
Mark Reid\altaffilmark{2} and Henrik Beuther\altaffilmark{3} }

\altaffiltext{1}{Max-Planck-Institut f\"{u}r Radioastronomie, Auf dem
H\"{u}gel 69, 53121 Bonn, Germany\email{lzapata,
kmenten@mipfr-bonn.mpg.de}}
\altaffiltext{2}{Harvard-Smithsonian
Center for Astrophysics, 60 Garden Street, MS-42, Cambridge, MA 02138
USA\email{reid@cfa.harvard.edu}}
\altaffiltext{3}{Max-Planck-Institut f\"{u}r Astronomie,
K\"onigstuhl 17, 69117 Heidelberg,
Germany\email{beuther@mpia-hd.mpg.de}}

\begin{abstract}
We present sensitive Very Large Array observations with an angular
resolution of a few arcseconds of the $J= 1 - 0$ line of SiO in the
$v$=1 and 2 vibrationally excited states toward a sample of 60
Galactic regions in which stars of high or intermediate mass are
currently forming and/or have recently formed.  We report the
detection of SiO maser emission in \textit{both} vibrationally excited
transitions toward only three very luminous regions: Orion-KL, W51N
and Sgr B2(M). Toward all three, SiO maser emission had previously
been reported, in Orion-KL in both lines, in W51N only in the $v=2$
line and in Sgr B2(M) only in the $v=1$ line.  Our work confirms that
SiO maser emission in star-forming regions is a rare phenomenon,
indeed, that requires special, probably extreme, physical and chemical
conditions not commonly found.  In addition to this SiO maser survey,
we also present images of the simultaneously observed 7 mm continuum
emission from a subset of our sample of star-forming regions where
such emission was detected.  This is in most cases likely to be
free-free emission from compact- and ultracompact-HII regions.
\end{abstract}


\section{INTRODUCTION -- HISTORY AND MOTIVATION}

SiO maser emission has been detected toward more than a thousand
asymptotic giant branch (AGB) stars and a few red supergiants
\citep[RSGs; see, e.g., ] []{Benson_etal1990, Habing1996}. These
objects have luminosities larger than a few thousand of \Lsun\ (AGB
stars) to $> 10^5$ \Lsun\ (RSGs), which are comparable to the
luminosities of intermediate- and high-mass protostellar and young
stellar objects.  Nevertheless, toward forming and young stars SiO
maser emission seems to be very rarely observable.

{\it Orion Kleinmann-Low (Orion-KL)} was the first source in the sky toward
which SiO maser emission was found (in the $v=1, J=2-1$ line) by
\citet{SnyderBuhl1974}. Subsequently, the first detections of the
$J=1-0$ line in the $v=1$ and 2 states, also toward, among others,
Orion-KL, were made by
\citet{Thaddeusetal1974} and by \citet{Buhletal1974},
respectively. Toward the Orion-KL region, maser emission was also found in
other SiO lines \citep[$v=1, J = 3 - 2$,][]{Davisetal1974}, including
transitions from the vibrational ground state
\citep{Tsuboietal1996}, but, at least to our knowledge, not from
states with $v > 2$, contrary to M-type stars.  Moreover, maser lines
from the $^{29}$SiO and $^{30}$SiO isotopomers
\citep{Olofssonetal1981}, and even the very rare $^{28}$Si$^{18}$O
were found \citep{Choetal2005}. Never detected, however, was the
$v=2, J= 2 - 1$ line.  The latter line also remains undetected in
oxygen-rich (M-type) evolved stars, although it is found in S-type
stars (in which O and C abundances are equal). This behavior is
explainable by the line's pumping mechanism \citep{Olofssonetal1981}.
Since the discovery of SiO maser emission in the Orion-KL region
\citep{SnyderBuhl1974}, only two more high mass star-forming regions,
W51N (= W51-IRS2) and Sgr B2(M)
\citep{Hasegawaetal1986, Ukitaetal1987} have been found to show
this kind of emission.

 In the past, a number of surveys have been undertaken with the goal
 of finding SiO masers in a larger number of star-forming regions.
 The number of sources surveyed for which upper limits have been
 published is less than two dozen and these upper limits are in the
 several Jy range \citep{Jewelletal1985, BarvainisClemens1984}, barely
 sensitive enough to potentially detect the W51N maser, and much too
 shallow to detect the $\sim1$ Jansky-strength maser in Sgr B2(M)
 \citep{Hasegawaetal1986, Ukitaetal1987}.

Around luminous AGB stars/RSGs, SiO maser emission arises from a
region of density around a few times $10^8$ and $10^9$ \ccm\ and
temperature above 1000 K, which is within a few stellar radii of
the photosphere \citep{Lockettetal1992,Bujarrabal1994}.  Given these
extreme requirements, one expects this emission to pinpoint in
star-forming regions the \textit{exact} location of the embedded
high-mass protostar that is exciting it, which frequently is not easy
to determine by other means
\citep[see discussion in][]{MentenReid1995}.
This is certainly borne out by the best-studied SiO maser associated
with a star formation region, that in Orion-KL.
\citet{MentenReid1995}, using simultanenous
high resolution VLA observations of the 43.2 GHz Orion-KL SiO maser
and weak continuum emission (from source-I), which they accurately
register with a 3.8 $\mu$m speckle image, clearly showed that at the
position of ``source-I'' no infrared source is detected. Thus, all
that is observed at infrared wavelengths ({\it i.e}. the famous IRc2) is
reprocessed radiation, while the position of the continuum emission,
which almost certainly comes from an ionized disk surrounding the
protostar, is right at the center of the SiO maser distribution
\citep{Reidetal2007}.  While source-I clearly is self-luminous, 
as argued by \citet{Reidetal2007}, it is unlikely to contribute a
significant faction of $\sim 10^5$ \Lsun\ of the KL region.

The SiO maser in W51N seems to be different from that in Orion-KL,
since it only appeared to show maser emission in the 42.8 GHz $v =2,
J= 1 - 0$ line, while that in Orion-KL shows emission in the $v=1$ and $2$
lines at comparable intensities. \citet{Moritaetal1992} located the
W51N SiO maser in a dense compact molecular core mapped in NH$_3$
emission by \citet{Ho_etal1983} and
\citet{ZhangHo1997}.
Very Long Baseline Array (VLBA) radio observations
by \citet{Eisneretal2002} reveal a comparable linear extent to that
observed in Orion-KL. It is clear from the latter data that, while
H$_2$O masers (also imaged by Eisner et al. 2002) trace a highly luminous
region on a few arcsecond (few tenths of a parsec) scales, only the
SiO maser marks, as in Orion, the exact location of a self-luminous
power source.

Sub-milliarcsecond resolution VLBA observations show that the W51N SiO
masers may be tracing the limbs of an accelerating bipolar outflow
close to the ``{\it dominant center}'' (where are located most of the H$_2$O, 
OH and SiO masers) \citep{Eisneretal2002}.  No radio continuum emission is detected toward
this SiO maser. However, the radio emission from Source-I would be
completely undetectable at the distance of W51 ($\sim$ 7 kpc).

Finally, little is known about the Sgr B2(M) SiO maser, from which
only the $v=1$ line had been found, except for the fact that it is
located close to (but not coincident with) the radio source F, which at
high resolution splits into several compact sub-sources
\citep{Gaumeetal1995}.  As in  W51N, the SiO maser is close to a
compact clump of \nhhh\ emission imaged by \citet{Vogeletal1987}.

In this work, we present a sensitive search for SiO maser emission
toward a sample of 60 high-mass star-forming regions using the Very
Large Array.  We report the detection of such emission in
\textit{both} vibrationally exited transitions ($v$=1 and 2) and
\textit{only} toward Orion-KL, W51N, and Sgr B2(M). This suggests that
SiO maser emission  is indeed a very special physical phenomenon
possibly only occurring within a short time period during the
formation of high-mass stars. Simultaneously, we obtained moderate
sensitivity data of those sources at 7 mm.

 In \S \ref{sample} we introduce our sample and in \S
 \ref{observations} discuss the observations undertook in this study.  
 In \S \ref{results} we present and discuss our
 SiO data and also our 7 mm continuum data.

 \section{\label{sample} THE SAMPLE}

 The 60 massive star-forming regions observed were selected from the
 sample of 69 high-mass protostellar objects of
 \citet{Sridharanetal2002} and the classical sample of mid-infrared
 selected star-forming regions studied by \citet{Willneretal1982}.
 The luminosities of these sources mostly range from several times
 $10^3$ to a few times $10^5$ \Lsun, which is comparable to the range
 from Mira variable to the RSG that are commonly
 found in the host stars of strong SiO masers.  We note that the
 luminosities of the SiO masers in the supergiants, {\it i. e.} VX Sgr or VY
 CMa, are typically 1 or 2 orders higher than those of Miras and
 comparable or even higher than those of the Orion-KL SiO maser.
 In Table \ref{table1}, we summarize the main properties of the 60
 objects selected.

\section{\label{observations}OBSERVATIONS}

 The observations were made with the NRAO\footnote{The National Radio
 Astronomy Observatory is a facility of the National Science
 Foundation operated under cooperative agreement by Associated
 Universities, Inc.} Very Large Array (VLA) between 2003 April 14 and
 May 1.  Two intermediate frequency (IF) bands were employed: one
 detecting right circular polarization, centered at 42.820555 GHz, the
 rest frequency of the $J = 1-0, v=2$ $^{28}$SiO line; the other, 
detecting left circular polarization, was centered at 43.122039 GHz,
 the rest frequency of the $J = 1-0, v=1$ $^{28}$SiO line. These rest
 frequencies were calculated using the high accuracy molecular 
 constants determined by Molla et al. (1991).

Both IF bands of the VLA correlator were configured in line mode with
32 channels covering 12.5 MHz, which provided 391 kHz (2.72 km
s$^{-1}$) resolution.   

The integration time on each source was about 5
minutes.  At this epoch, the VLA was in its D configuration.  The
absolute flux density calibrator was 1331+305, for which we adopted a
flux density of 1.47 Jy.  Amplitudes and phases were calibrated by
observations of compact extragalactic radio sources close in position
to our program sources.  In Table \ref{caltable}, we present the list
of the compact extragalactic radio sources used as secondary
calibrators and their bootstrapped flux densities.
 
 The data were edited and calibrated in the standard manner using the
 software packages AIPS and MIRIAD developed by the NRAO and
 BIMA\footnote{The Berkeley Illinois Maryland Association}. Maps were
 obtained using the AIPS task IMAGR and the MIRIAD tasks INVERT,
 CLEAN, and RESTORE.  For sources free of line emission, continuum
 maps were produced from ``{\it channel 0}'' data, {\it i.e.}, pseudo-continuum
 ($u,v$)-databases containing the inner 75\% of the bandpass. For the
 sources showing maser emission, the ($u,v$)-data channels free of
 line emission and not affected by band edge were averaged and imaged.
 For most of the 7 mm continuum maps, we used the ROBUST weighting
 parameter set to 0, for an optimal compromise between sensitivity and
 angular resolution. However, for fainter sources we use ROBUST=5,
 which corresponds to natural weighting, to achieve maximum
 sensitivity in each continuum image. Finally, the strongest 7 mm
 continuum sources were self-calibrated in phase and amplitude. The
 resulting {\it rms} noise of the line images was better than about 10 mJy
 beam$^{-1}$ per channel. In a few sources, it was higher due to
 unfavorable weather conditions and/or low declinations. The results
 of the continuum imaging are discussed in \S\ref{contresults}.

\section{\label{results}RESULTS AND DISCUSSION}

\subsection{The SiO Maser Emission}

 In Table \ref{Params}, we present the main results of our search for
 SiO masers in both vibrationally excited ($v$=1 and 2) transitions
 toward the 60 high- and intermediate-mass star forming regions. 
 We detected SiO $J= 1 - 0$ maser
 emission in both the $v=1 $ and 2 lines above the 5-$\sigma$ levels
 (or about 50 mJy) only toward Orion-KL, W51N, and Sgr B2(M).

\subsection{Source-I}
 The SiO maser emission from both vibrationally excited transitions
 associated with Orion-KL shows its characteristic double-peak profile
 centered near 6 km s$^{-1}$ (the systemic velocity of the molecular
 cloud core, see Figure \ref{f1}).  As already found in many previous
 studies dating back to \citet{SnyderBuhl1974}, two groups of features
 are observed, in the $[-14,+3]$ \kms\ and $[+8,23]$ \kms\ intervals,
 with little emission in between.

 It is interesting to note that the flux density of the SiO maser
 emission toward the sources Sgr B2(M) and Orion-KL in the $v$=1
 vibrationally excited transition is stronger than that in the $v$=2
 transition, while in W51N the reverse is true (see Figure \ref{f1}).

 Finally, comparing the flux densities of the SiO maser emission 
 in the Orion-KL, W51N and Sgr B2(M), we have calculated that the SiO masers 
 in Orion-KL are approximately one order of magnitude more luminous than those in W51N
 and Sgr B2(M) (see Table 3).

\subsubsection{The Nature of the SiO Emission}

 The SiO maser emission toward high-mass star forming regions has been
 associated with powerful bipolar protostellar outflows
 \citep{Greenhilletal1998, Eisneretal2002}.  
 VLBA radio observations (with $\sim1$ mas resolution) of the Orion-KL SiO
 masers were consistent with an origin in a wide-angle biconical flow
 (with a ``X''-like pattern) with the SiO masers outlining the limbs of
 the outflow cavity and centered on the source-I
 \citep{Greenhilletal1998}.  More recent data, however, suggests that
 the SiO masers originate in the material expelled from a rotating disk
 around ``source I'' \citep{Greenhilletal2004,Reidetal2007}.

 \subsection{W51N, Sgr B2(M), and Source-I}

 We found SiO $J = 1 - 0$ maser emission in \textit{both}
 vibrationally excited transitions toward W51N and Sgr B2(M) (see
 Figure \ref{f1}).

 Toward Sgr B2(M)
 the $v$=1 SiO maser emission shows a single peak that is redshifted
 from the cloud systemic velocity (about 60 km s$^{-1}$), 
 has a linewidth of about 13 \kms and is
 centered at a v$_{LSR}$ of 87 km s$^{-1}$.  The $v$=2 transition only shows 
 one redshifted very 
 narrow line centered at a v$_{LSR}$ of 96 km s$^{-1}$.  
 The latter line is unresolved with our 2.72 \kms\
 resolution, and is atypically narrow for an SiO maser line. We are
 nevertheless convinced of its reality as the emission arises from the
 same position as the $v=1$ emission.

 Toward W51N the $v$=1 SiO maser line is spread over 12 \kms\ and is
 centered near v$_{LSR}$ = 50~km~s$^{-1}$, close to the systemic velocity 
 of the molecular cloud core associated with W51.  The $v$=2 transition shows
 emission over 20 \kms, centered at a similar velocity of v$_{LSR}$ =
 49 \kms.  Neither lines exhibit a double peak
 appearance.

 In the W51N region, the SiO maser emission is tracing a high velocity
 ($\sim$ 80 km s$^{-1}$) bipolar protostellar outflow with a size of
 about 10$^4$ AU that emanates from a molecular core observed in NH$_3$
 (Morita et al. 1992) as discussed before.  We note that in the case of
 Sgr B2(M) the SiO maser emission only shows high velocity redshifted emission
 ($\sim$ 30 km s$^{-1}$) emanating from a molecular NH$_3$ core 
 (Morita et al. 1992). 

\subsubsection{Variability}

Toward two of our SiO maser associated with SFRs (Source-I and W51N)
we see clear signs of variability, but only for Source-I do extensive
monitoring data exist.  Comparing the flux density of the SiO $v = 1,
J = 1-0$ transition reported by Menten and Reid (1995) toward
Source-I with that measured by us (Figure \ref{f1} and Table \ref{Masers}) we found
a $\sim3$ times lower value.

Long term monitoring of the Orion-KL $J=1-0$ SiO maser lines generally
shows the $v=1$ line stronger then the $v=2$ line
\citep{Martinezetal1988, Alcoleaetal1999, Pardoetal2004} with the ratio
varying between $>4$ and $<2$. The ratio we measure ($\sim2$) is in
line with this behavior and our flux densities are at the low end of
the values found during the monitoring.

\subsubsection{Comparison with Red Giant and Supergiant Stars}

The long timerange monitoring of SiO $J=1-0$ masers with the Yebes 14
m telescope
\citep{Martinezetal1988, Pardoetal2004} reveals
that both the \textit{v=1} and the \textit{v=2} lines exhibit dramatic
variability in some Mira stars. In $o$ Ceti, for example, the SiO
luminosity may vary by a factor $> 500$ over a 332 d variability
cycle, with a maximum SiO luminosity near the same time as the
infrared radiation. Higher luminosity/higher mass-loss RSGs show longer 
term SiO variability and higher SiO maser luminosities.
Both SiO lines toward the supergiants VY CMa and VX Sgr show a long-term
decline with a drop to half intensity over $~\sim 1500$ days, longer
than the infrared variability cycles of these objects.

Towards $\mu$ Cep the intensity in both lines increased $>5$ fold over
an 800 day period and then declined by an even larger factor over a
similar interval and has been at a low level for years.  Finally, a
number of well-known RSGs (e.g. $\alpha$ Ori, $\alpha$ Sco, and
$\alpha$ Her) do no show any maser emission at all.

It is interesting to compare the luminosities of the SiO masers in SFR with
the typical luminosity ranges of Mira 
and RSG SiO masers that we have
added to Table \ref{Masers}. It looks like as if the
SFR SiO maser luminosities are higher than the luminosities of Mira
stars but lower than that of the RSGs. The same appears to be true for
the luminosities of the exciting stars.

\subsection{\label{contresults}7 mm Continuum Emission}

Of the 60 massive star-forming regions observed in our survey, 7 mm
continuum emission was only detected toward 17 above a 4-$\sigma$
level. Continuum maps of these sources are shown in Figure \ref{f1} 
and flux densities and source sizes are given in Table 4.
The continuum emission is most likely free-free emission from compact
and ultracompact HII regions.  However in a few sources, there are
indications that it might originate from ionized jets, dusty envelopes
and/or disks.  In Table 5 we give a tentative interpretation for the
nature of these sources, based mostly on other centimeter/millimeter
continuum observations at high angular resolution and, in some cases, on their
spectral energy distributions (SEDs).

We detected 14 ultracompact and compact H II regions: N6334I, Sgr
B2(M), AFGL 2046, AFGL 2136, W51N, W51e1, W51e2, W51IRS1, AFGL 2591,
NGC7538 IRS1-A, NGC7538-IRS1-B, W3 IRS5, W3(OH), and MonR2. For all of
these sources, data taken at other wavelength have been reported
earlier (see Table 5).  The sources show three different type of
morphologies: cometary, unresolved, or irregular (see Figure
\ref{f1}), with the cometary morphology dominating.

It is interesting to note that out of 32 objects selected from the
sample of high-mass protostellar objects of \citet{Sridharanetal2002},
only one was detected at 7 mm (I18440$-$0148).  This could be
explained if these objects are young (proto)stars with no or very
faint free-free emission (less than a few mJy) as selected by
\citet{Sridharanetal2002}. Dust emission at 7 mm wavelength is
expected to be very faint and not detectable with our integration
times. Moreover, \citet{Zapataetal2006} using observations of the VLA found that
the 7 mm continuum emission from 10 massive dusty objects of this
sample is less than a few mJy.

At a 4-$\sigma$ level ($>2$ mJy), we do not detect the arcsec-sized 
hypercompact HII regions located in the NGC7538 IRS9, W33A, and AFGL 2591  
and that one reported by \citet{vanderTakMenten2005} toward W3IRS5.
Neither do we detect source-I in Orion, the
continuum source associated with the SiO masers
\citep{MentenReid1995, Reidetal2007}. 
Note, however, that the noise level in our Orion continuum image is
particularly high, mainly as our bandwidth provided not enough channels free of line emission
to make the continuum image. Also, we do detect the hypercompact HII region RS4
discussed by \citet{MentenvanderTak2004}.  Finally, we note that the
continuum flux density at this wavelength for W3(OH) is low compared
to those fluxes reported at 1.3 cm and 3 mm by
\citet{Wilneretal1995,Wilsonetal1991}.   
It could be explained if the source is varible in time as reported at 1.3 cm by
\citet{Guilloteauetal1993}. 

\subsubsection{Comments on Selected Individual Sources}

\subsubsubsection{Sgr B2(M)}

As shown in Figure \ref{f2}, we found two strong 7 mm continuum
sources in Sgr B2(M) (A in the west and B in the east). These
two sources are the counterpart of the compact objects F1--F4 
and the compact HII region ``I'' reported by \citet{Gaumeetal1995} 
at 1.3 cm, but due to our poor angular resolution ($\sim$ $5''$ $\times$ $1''$ 
at P.A.=150$^\circ$) we could not resolved them.

The F1--F4 objects have been already imaged at 7 mm 
with much higher angular resolution (0$\rlap.{''}$06), 
with the VLA \citep{DePreeetal1998} and were resolved into a total of
$\sim20$ separate HC HII regions.

\citet{Kuanetal1996} have compiled continuum flux densities for Sgr B2(M)
at frequencies between 4.9 to 8.3 GHz and between 78.5 to 109.9 GHz.  Our
43 GHz observations bridge an important frequency gap, and the total
flux density we measure for Sgr B2(M) indicates a fairly flat spectrum
with dust only contributing significantly at higher frequencies. While
the numerous UC HII regions present in Sgr B2(M) are expected to have
rising or falling spectra, the effective spectral index is likely the
result of more extended (flatter spectral index) emission dominating
the integrated emission analyzed by \citet{Kuanetal1996} and us.

\subsubsubsection{I18440$-$0148}

{I18440$-$0148 was the only one from the sample of
\citet{Sridharanetal2002} detected at 7 mm. From the 3.6 cm continuum
flux $\leq$ 1 mJy
\citep{Sridharanetal2002} and the 7~mm flux measured by us (see Table~4), we calculated a spectral
index of $\alpha \geq 1.1$.  We, thus, suggest that the 7 mm continuum
emission of this source is moderately optically thick free-free
emission from an ultra-compact HII region.

\subsubsubsection{DR21(OH)}

DR21(OH) (or W75S) is a very young high mass star forming region with
a bolometric luminosity of $1\times10^3$~\Lsun\ showing very faint
centimeter continuum emission \citep{argonetal2000}.  It also shows
strong compact millimeter continuum emission \citep{Woodyetal1989,Padinetal1989} 
and H$_2$O, OH and class I as well as class II CH$_3$OH maser emission
\citep{argonetal2000,Magnunetal1992,Plambecketal1990,Menten1991}. Class
I CH$_3$OH masers trace a powerful east-west bipolar outflow almost
perpendicular to the plane of sky \citet{koganetal1998}. Toward this
region we detect a compact source that is the 7 mm counterpart of the
millimeter source MM1 \citep[][]{Woodyetal1989,Padinetal1989}.  
This source is found to be resolved at this wavelength (see Table 4).

From the 3.6 cm continuum peak flux density value ($\sim$ 0.31 mJy)
reported by \citet{argonetal2000} of this region and our 7 mm value
(Table 4), we calculate a spectral index of $\alpha$=1.98$\pm$0.2 that
suggests that its cm continuum emission maybe optically-thick free-free
emission from a ultra-compact HII region.  However, if we take the 2.7
mm continuum flux density ($\sim$ 0.3 Jy) reported by
\citet{Woodyetal1989} for MM1, we find a much steeper spectral index
of $\alpha$=3.6$\pm$0.4, very close to that derived by
\citet{Woodyetal1989} from the 1.4 and 2.7 mm continuum measurements.  
We think that the emission at 7 mm might be partially optically thin dust emission  
from either a dusty envelope or disk, or both.

\subsubsubsection{AFGL 490}

The AFGL 490 region is located at about 1 kpc distance
\citep{Snelletal1984}, has a mass of 490 \Msun, a bolometric
luminosity of $2 \times 10^3$ \Lsun\ and is still embedded in its
parent molecular cloud 
\citep[A$_V$ $\sim$ 40 mag,][]{Alonso-Costa1989}. Its central object
drives a high-velocity outflow \citep{Mitchelletal1995}.  Very
recently \citet{Schreyeretal2006} by using $^{17}$CO line and 1.3 mm
continuum observations with an angular resolution of about 1$''$ found
evidence of rotating dusty disk driving the molecular outflow.

In this region we found a compact ($\sim$ 2000 AU size) source that is
the 7 mm counterpart of the millimeter source reported by
\citet{Schreyeretal2006}.

\subsubsubsection{AFGL 961}

AFGL 961 is also a relatively close-by high-mass star
forming region with a bolometric luminosity of about 10$^4$
\Lsun. This region is composed of two near-IR objects (E and W),
separated by about $5''$ (8000 AU). Castelaz et al. (1985) show that
the spectral energy distribution of GL 961-E dominates at wavelengths
longer than $2.2\ {\rm\mu m}$, while the western object is brighter at
shorter wavelengths.  This binary system is located at the center of a
parsec scale molecular outflow oriented in the N-S direction with the
approaching lobe towards the north (Lada \& Gautier 1982). Very
recently \citet{Alvarezetal2004} found a third faint infrared object
associated with the western source.

We found a faint and unresolved 7 mm continuum source that is the
millimeter counterpart of the infrared source GL 961-E.  Single-dish
(sub)millimeter 350, 870, and 1300 $\mu$m data all taken with beam
sizes of $30''$ or larger cannot discriminate between AFGL 961W and E
\citep{Guertleretal1991}. If we \textit{assume} that all the
(sub)millimeter flux and the radio emission arise from the same source
(AFGL 961E) we can construct the SED, shown in
Figure \ref{f4}. It shows a spectrum in which the emission is dominated
by a component that rises rapidly with frequency with a spectral index
of $\alpha=3.0$. A simple interpretation would be that the 7 mm
continuum emission from GL 961-E is associated with dust emission from
a core or disk. Clearly, however, more high resolution radio and
(sub)mm observations are needed to test this interpretation.

\section{SUMMARY}

We report the detection of SiO maser emission in \textit{both}
vibrationally excited transitions from massive star forming regions
only towards three  very luminous regions (Orion-KL, W51N and SgrB2(M)) out of the 60 observed galactic 
star-forming regions. Toward all three, SiO maser emission had previously been
reported, in Orion-KL in both lines, in W51N only in the $v=2$ line
and in Sgr B2(M) only in the $v=1$ line.  Since we do not detect
either SiO line in 57 other regions of intermediate- and high-mass
star formation, our work confirms that such emission in star-forming
regions is a rare phenomenon, indeed, that requires special,
probably extreme, physical and chemical conditions not commonly found.
In addition to this SiO data, we also present images of the
simultaneously observed 7 mm continuum emission from a subset of our
sample of star-forming regions where such emission was detected.  This
is in most cases likely to be free-free emission from compact- and
ultracompact-HII regions.

\acknowledgments
We thank the anonymous referee for many valuable suggestions.
This research has made extensive use of the SIMBAD database, operated
at CDS, Strasbourg, France, and NASA's Astrophysics Data System.

\bibliographystyle{apj}
\bibliography{luis}

\clearpage

\begin{figure}
\begin{center}
\includegraphics[angle=0,scale=1]{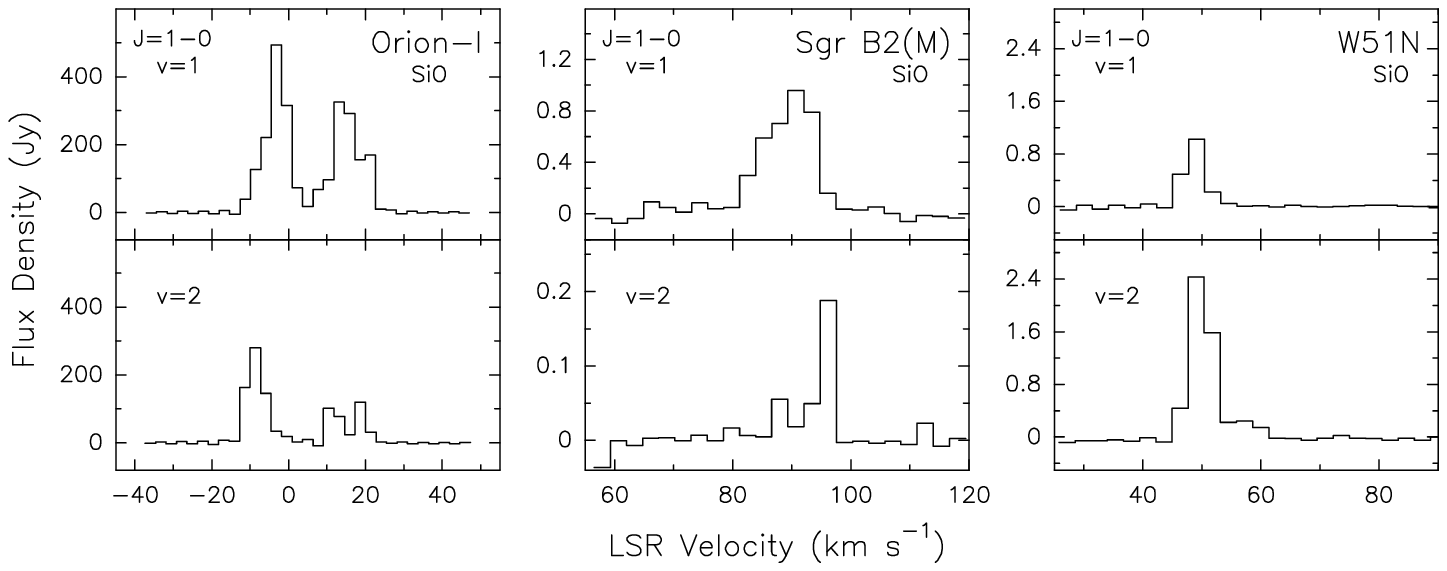}
\end{center}
\caption{\label{MaserSpectra}Spectra of the SiO $J=1-0$, $v$=1 lines (top) 
 and $v=2$ lines (bottom) observed toward Source-I (left), Sgr B2(M) (middle), and
 W51N (right).  The spectral resolution is  2.72 km s$^{-1}$.\label{f1}}
\end{figure}

\begin{figure}
\begin{center}
\includegraphics[width=54mm,height=54mm]{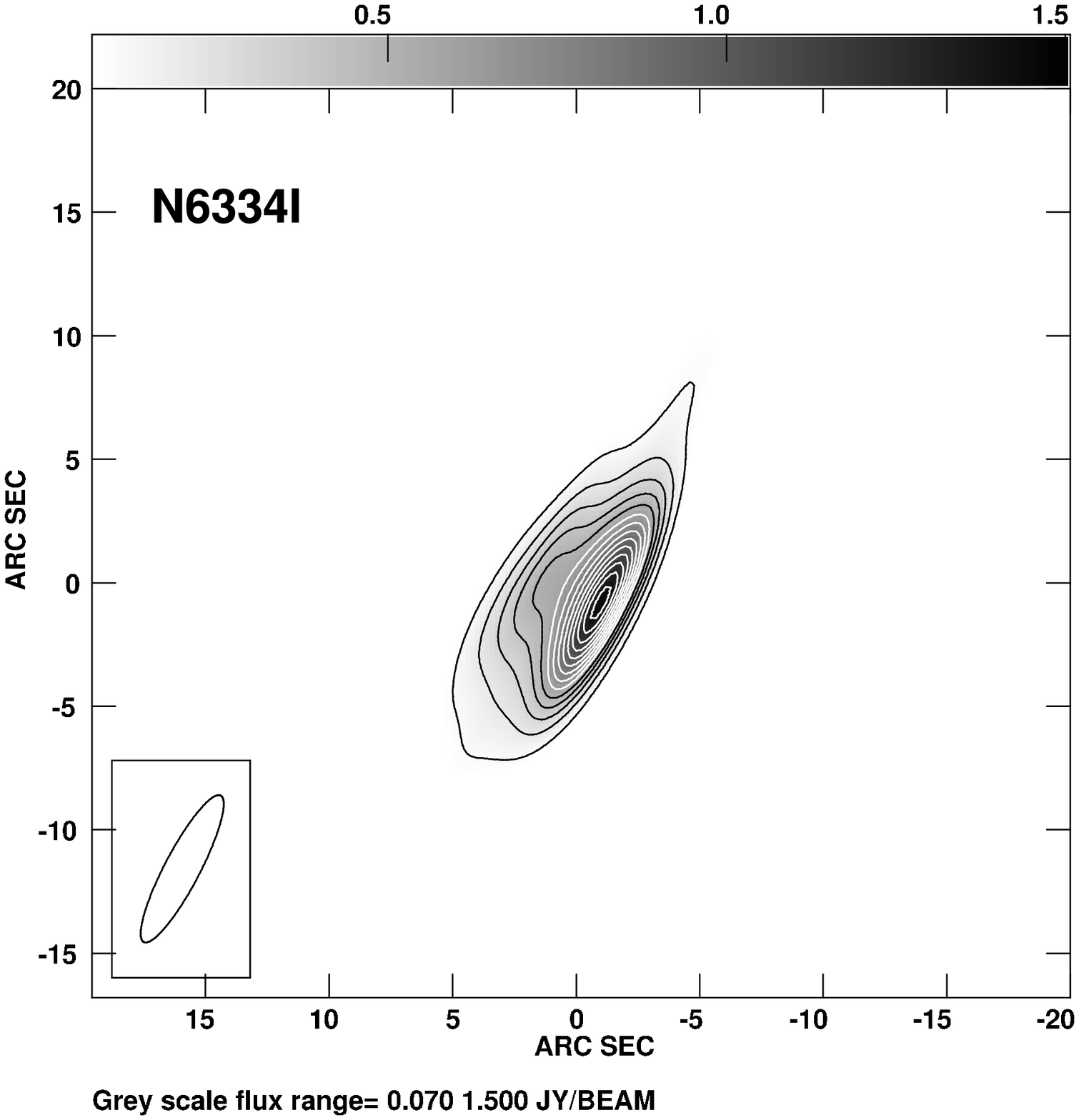}
\includegraphics[width=54mm,height=54mm]{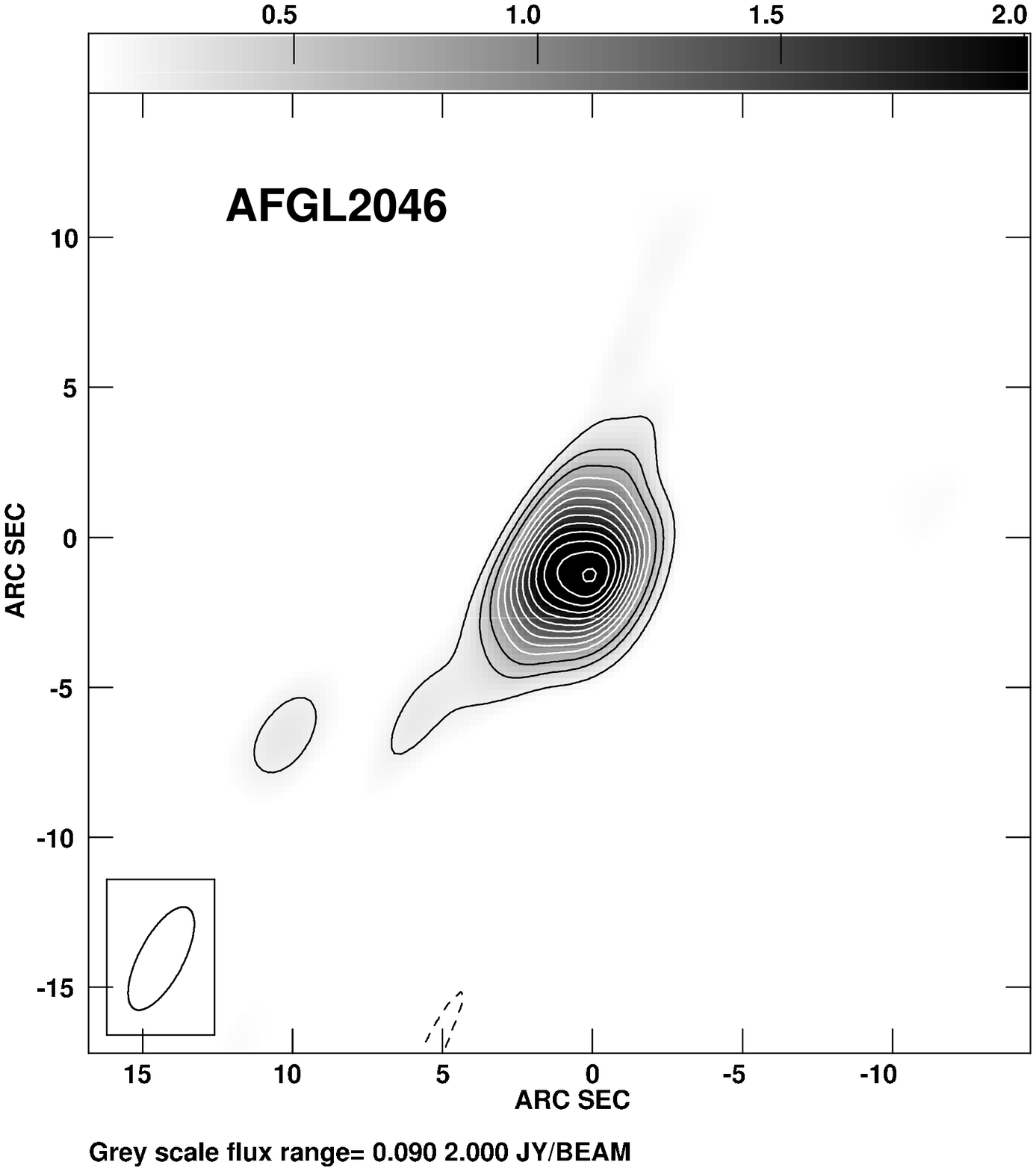}
\includegraphics[width=54mm,height=54mm]{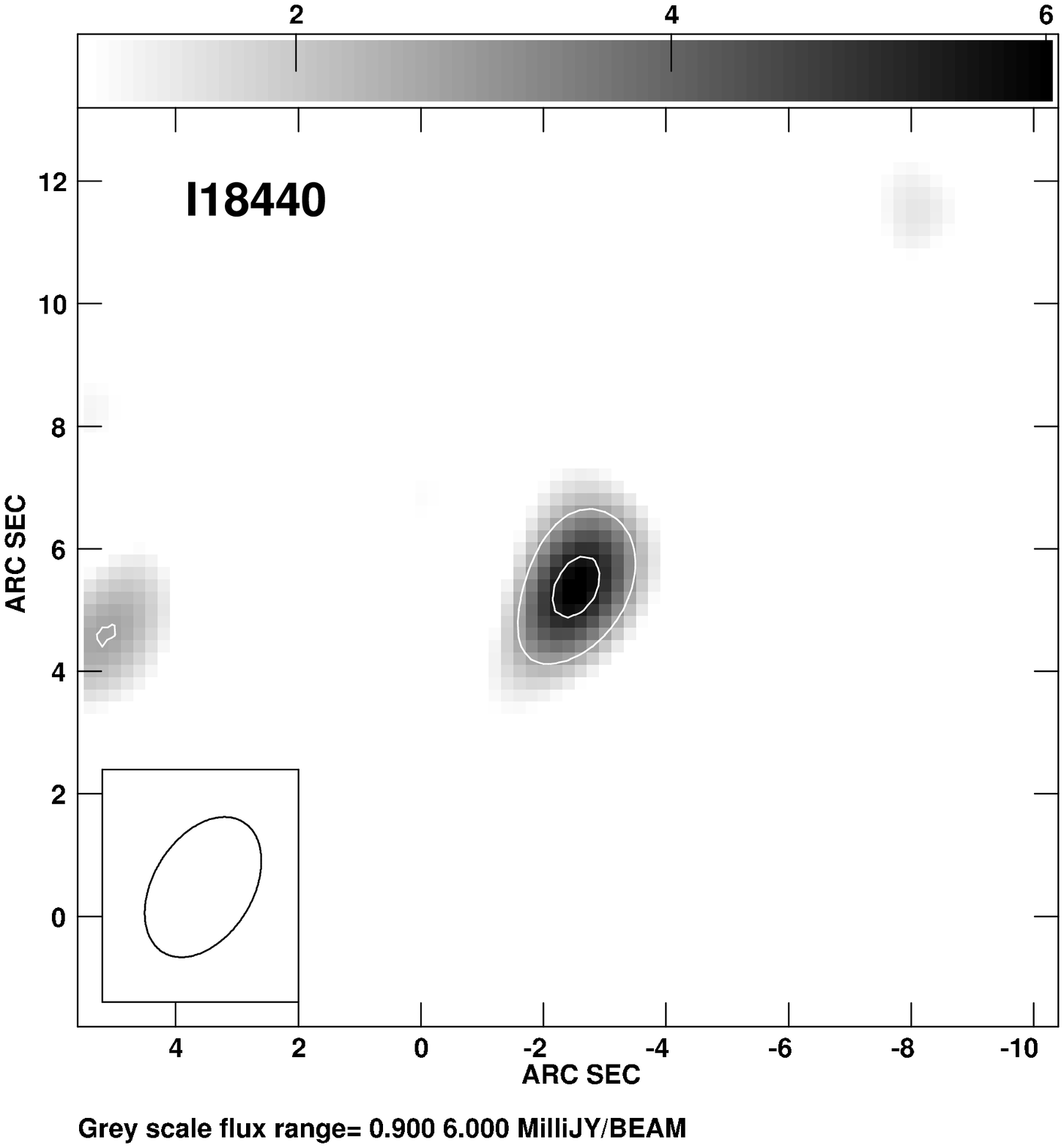}\\
\includegraphics[width=54mm,height=54mm]{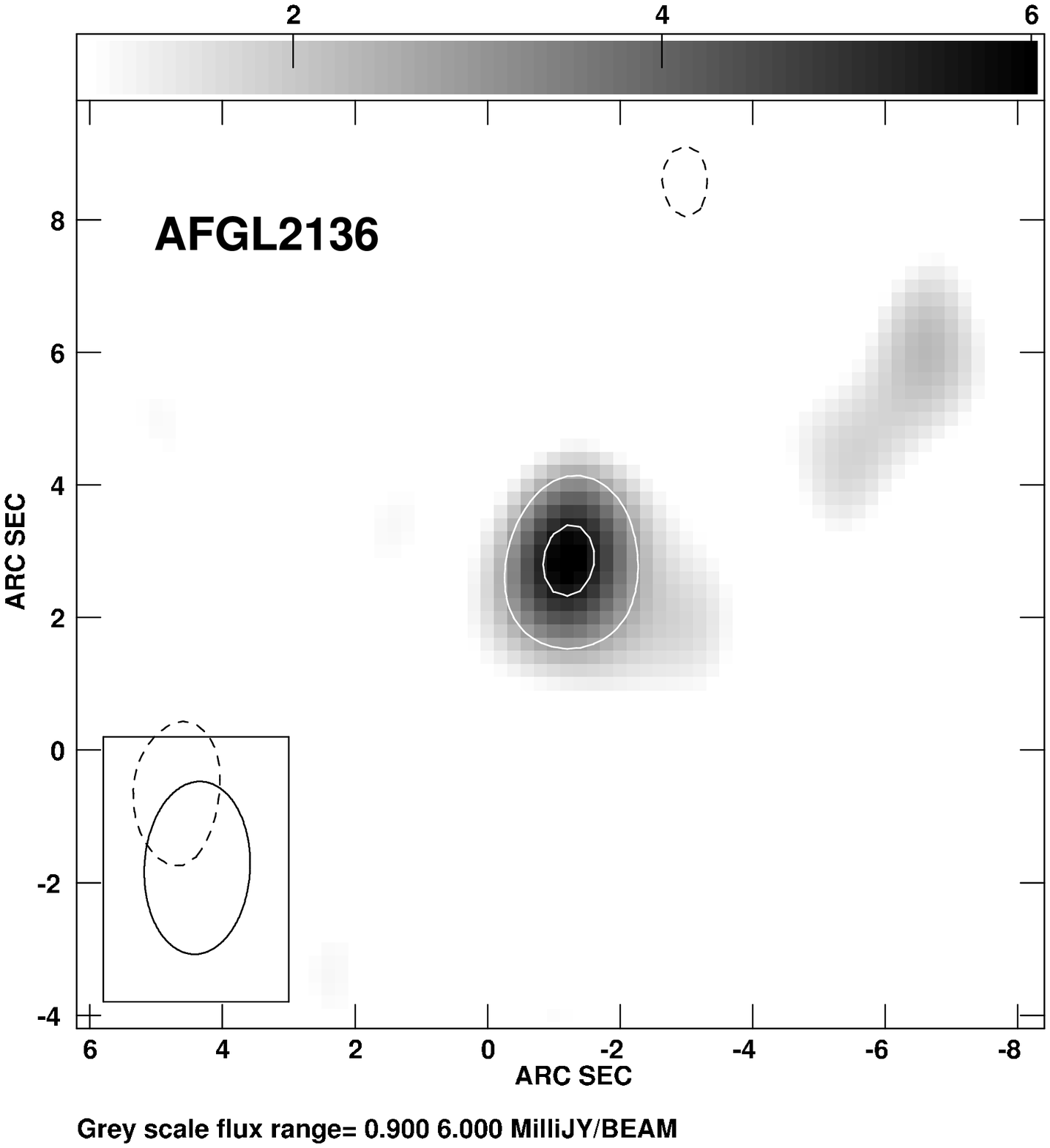}
\includegraphics[width=54mm,height=54mm]{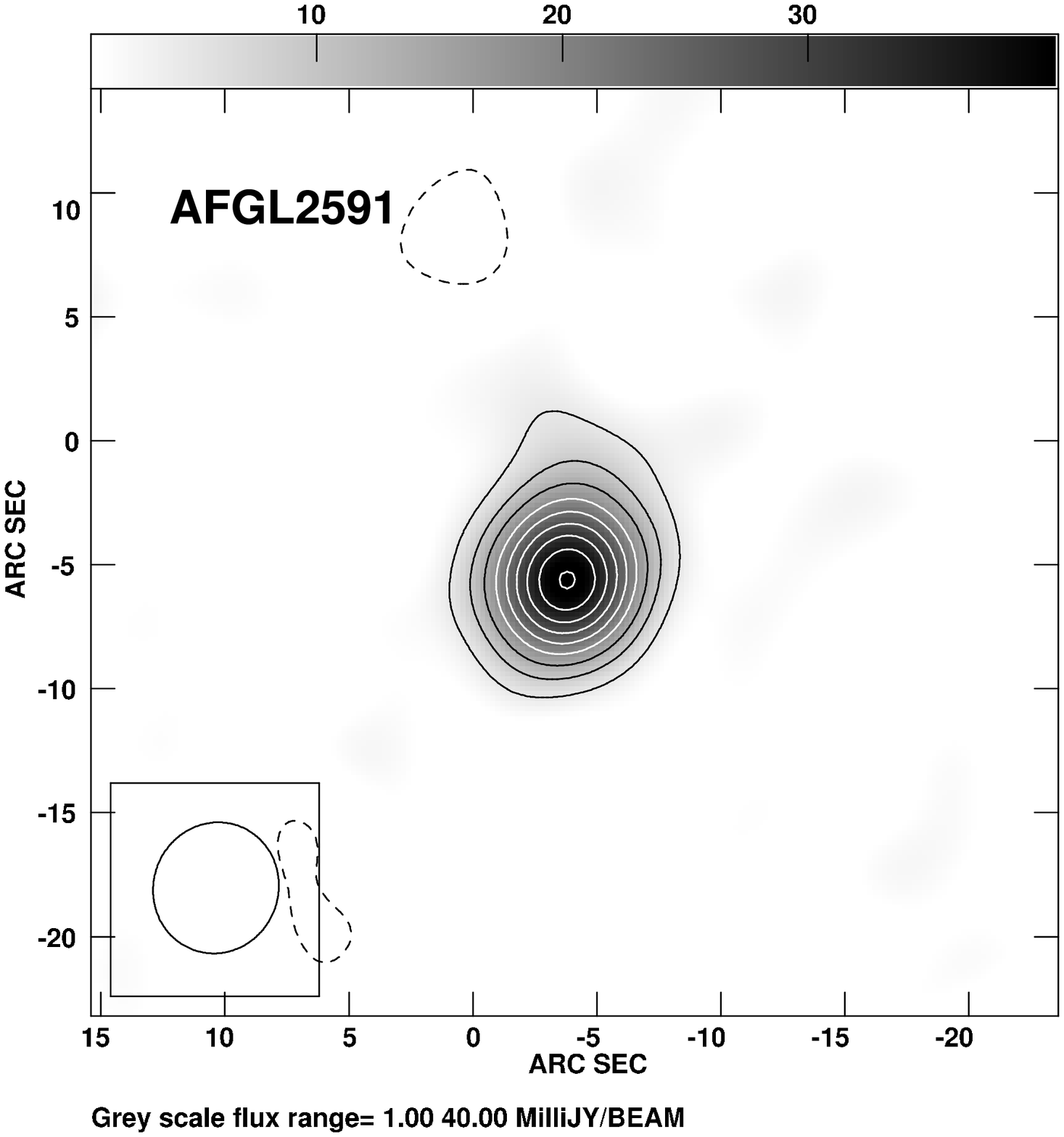}
\includegraphics[width=54mm,height=54mm]{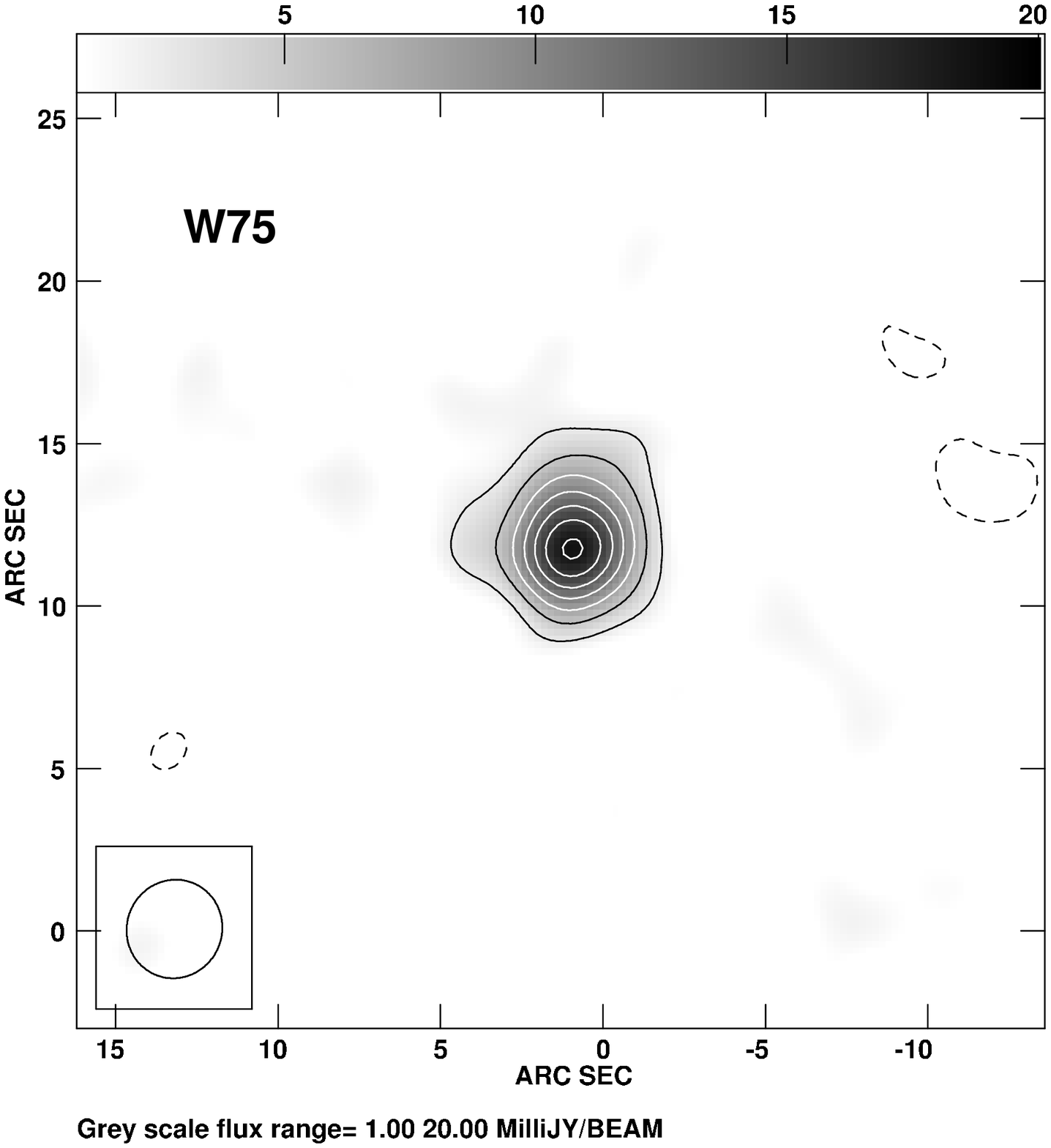}\\
\includegraphics[width=54mm,height=56mm]{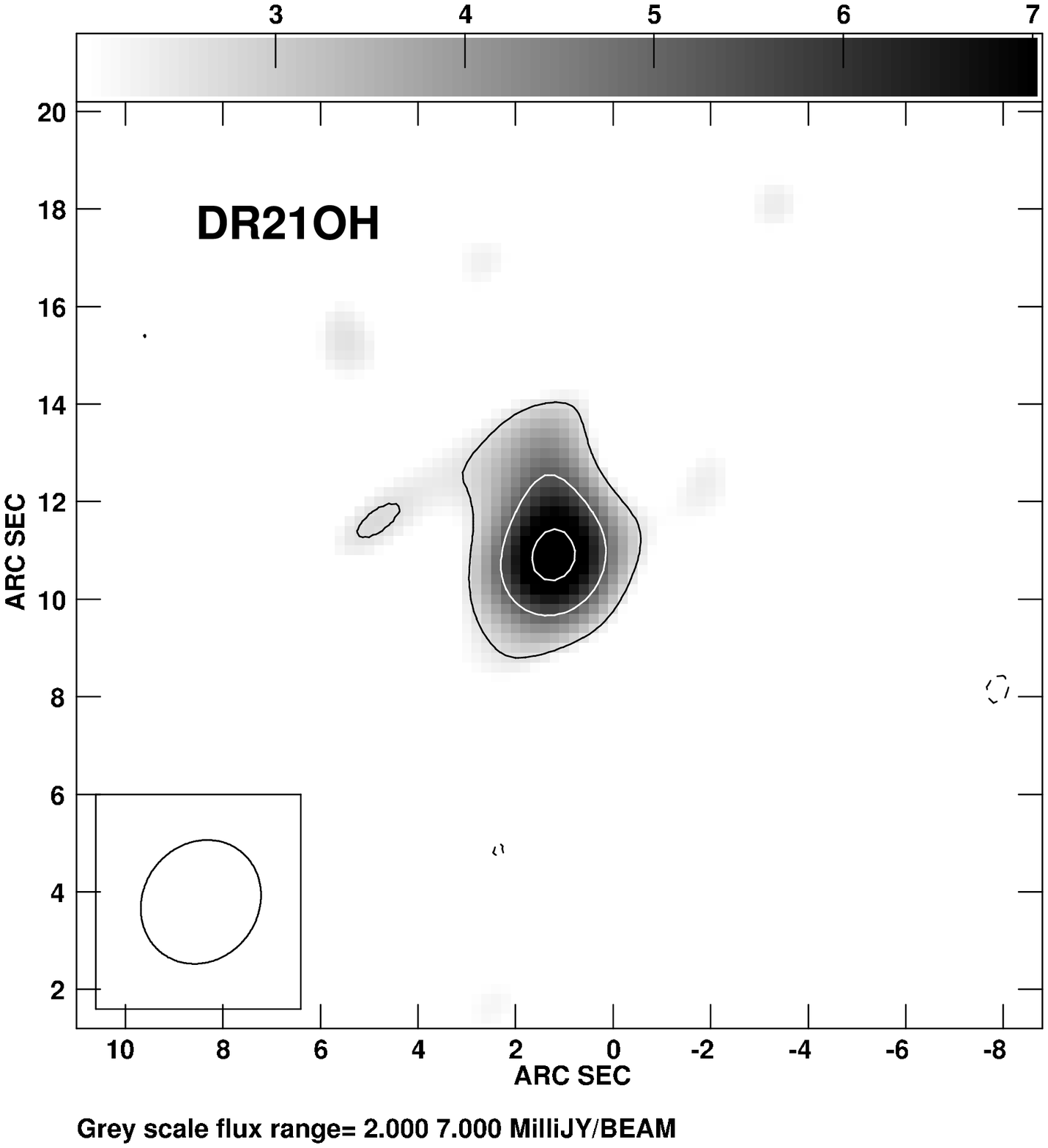}
\includegraphics[width=108mm,height=56mm]{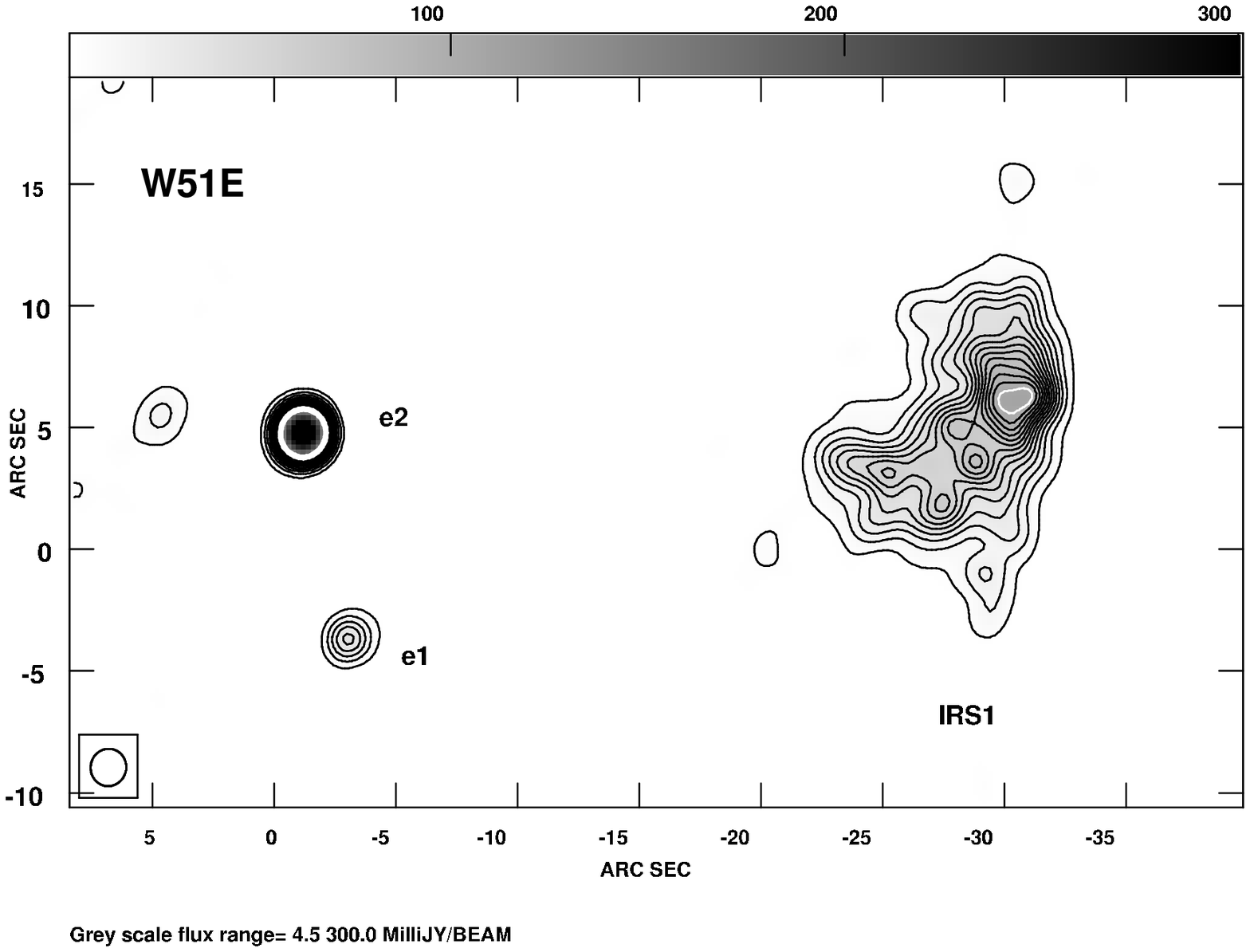}
\end{center}
\caption{\label{ContImages}VLA 7mm continuum images. The half-power 
contour of the synthesized beam
is shown in the bottom left corner of each image.  The contours for
the most of the sources are -5\%, 5\%, 10\%, 15\%, 20\%, 25\%, 30\%,
35\%, 40\%, 45\%, 50\%, 55\%, 60\%, 65\% 70\%, 75\%, 80\%, 90\% and
95\% the peak flux shown in Table 4.  
The var scale on the top of each image gives approximately the 7 mm peak continuum
emission, and the range and units are indicated in the bottom part. The x and y
coordinates in each sub-image are offsets (in arcsec) in right
ascension and declination direction, respectively, from the positions
listed in Table 1. \label{f2} The continuum images of Sgr B2(M) and
W51N were made by averaging only the free-line channels. The white and
black crosses indicate the positions of the SiO $J=1-0$ masers
associated with Sgr B2(M) and W51N, respectively.  }
\end{figure}

\clearpage
\addtocounter{figure}{-1}

\begin{figure}
\begin{center}
\includegraphics[width=54mm,height=54mm]{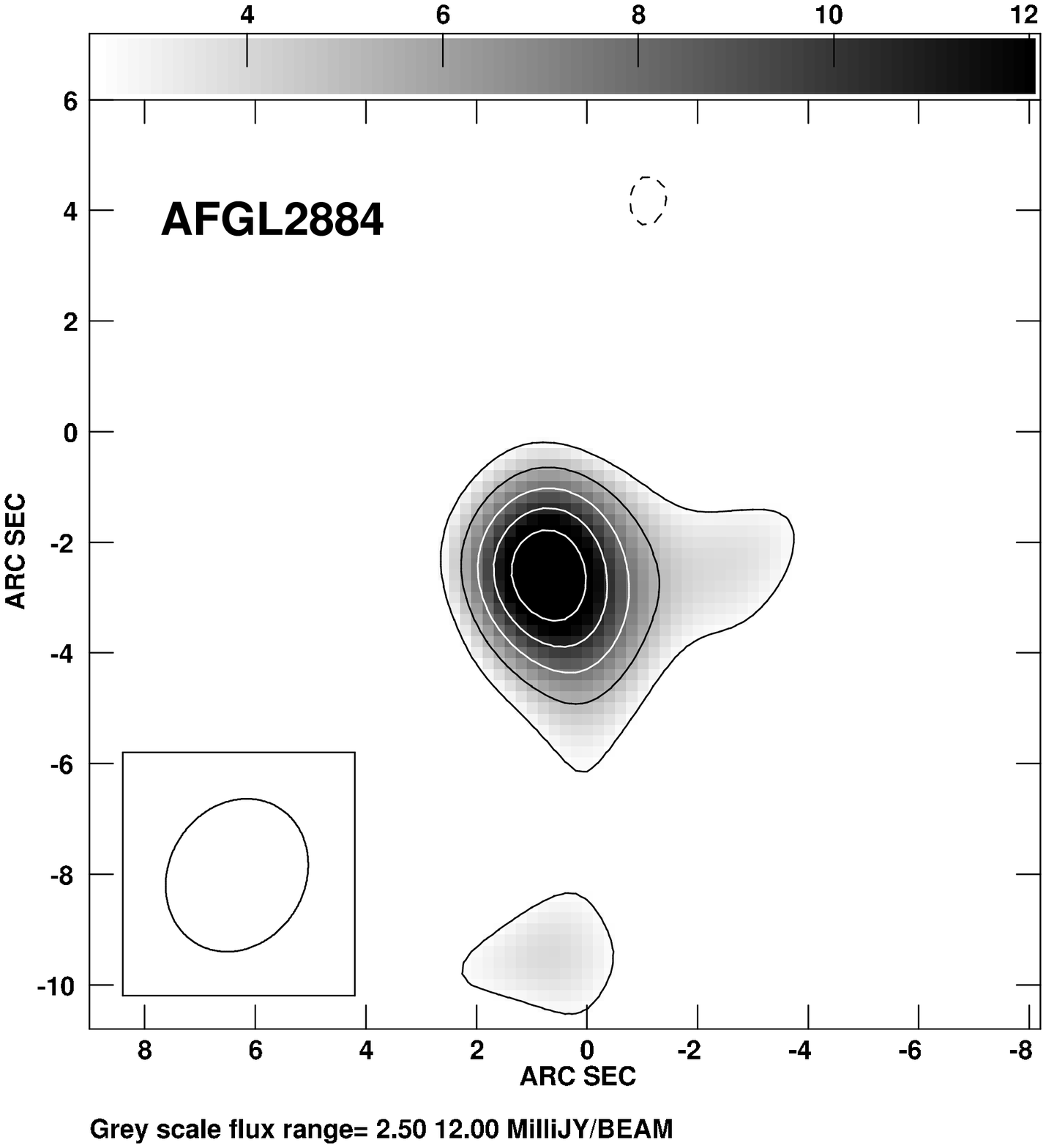}
\includegraphics[width=54mm,height=54mm]{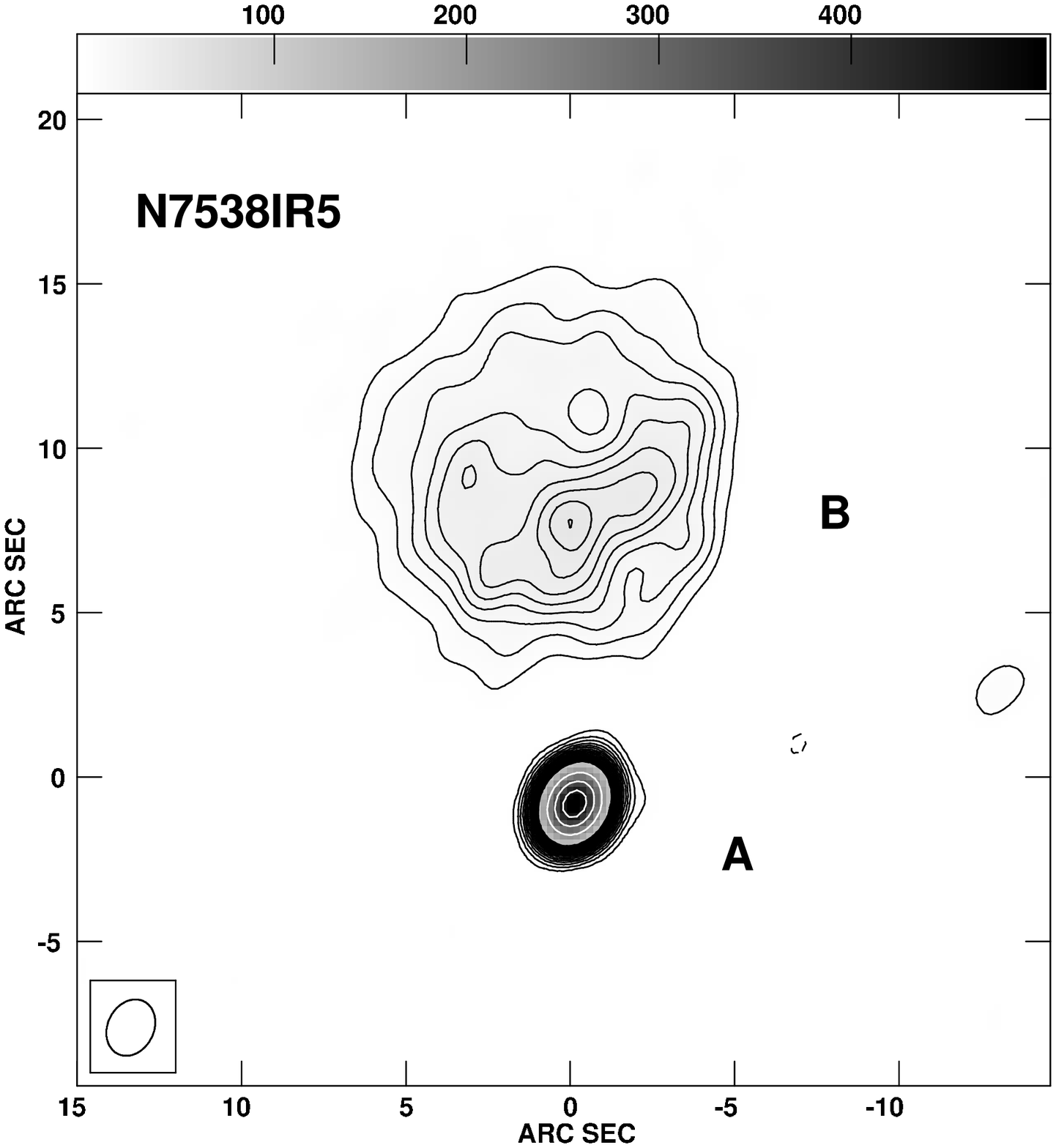}
\includegraphics[width=54mm,height=54mm]{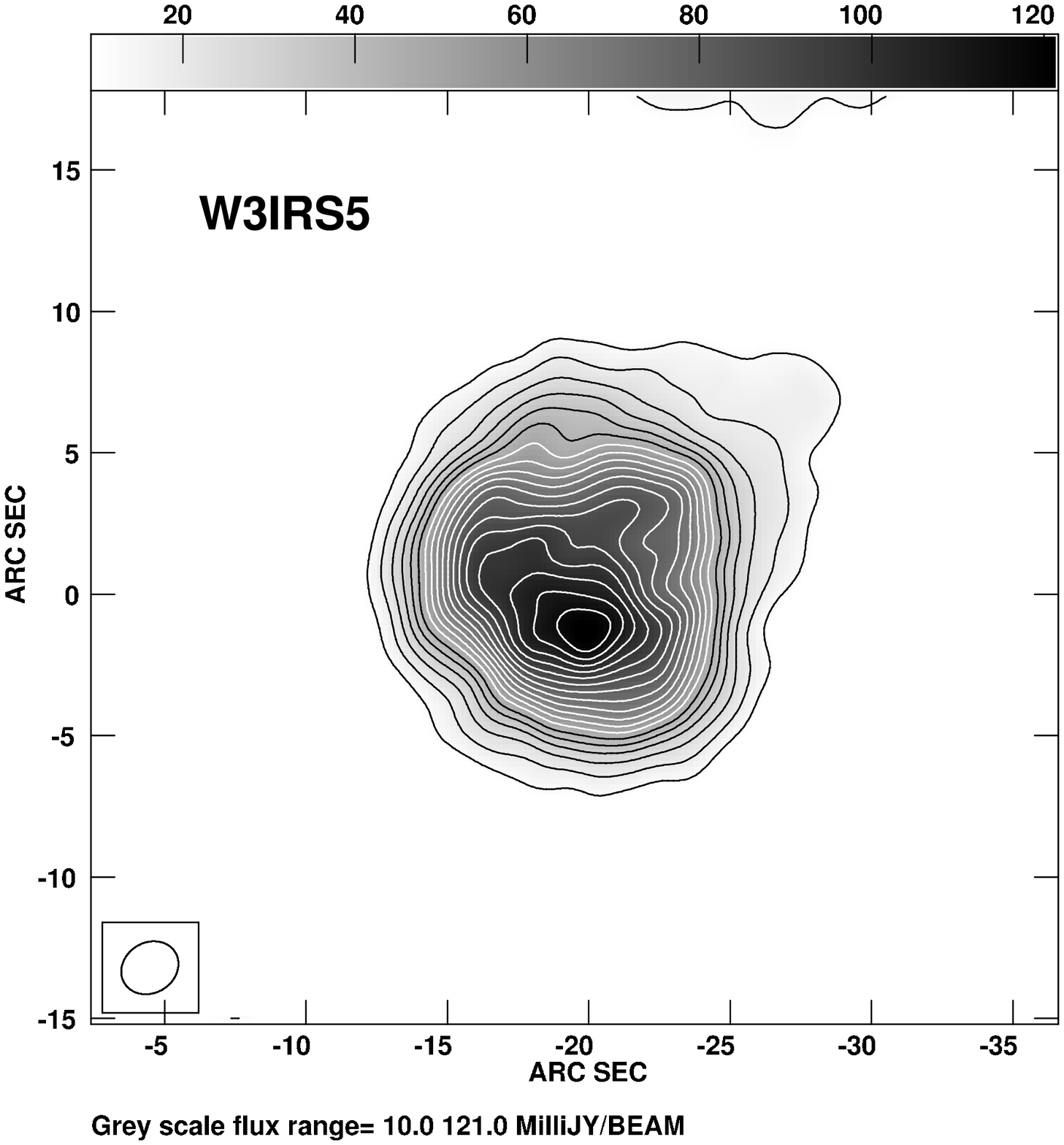}\\
\includegraphics[width=54mm,height=54mm]{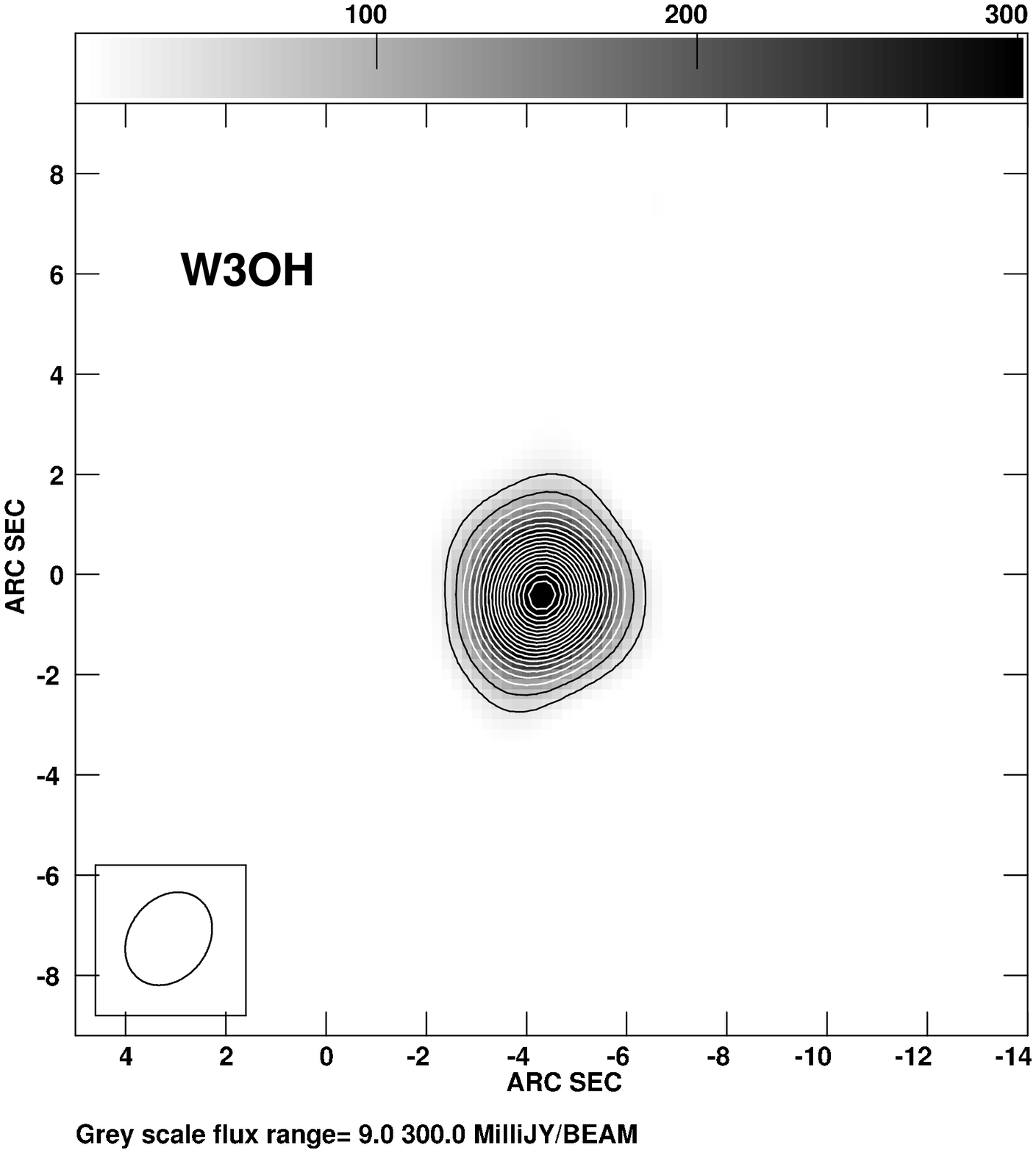}
\includegraphics[width=54mm,height=54mm]{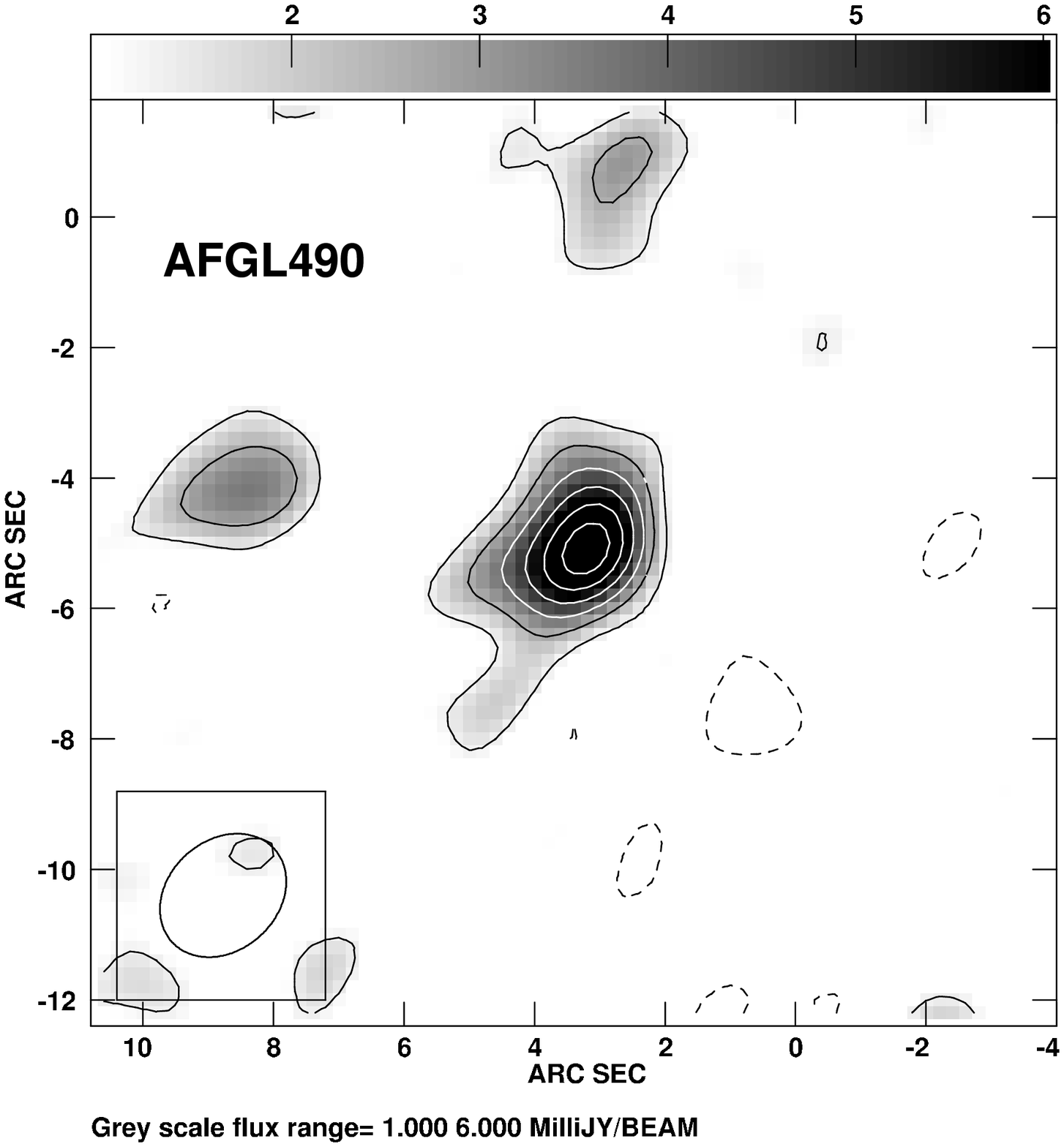}
\includegraphics[width=54mm,height=54mm]{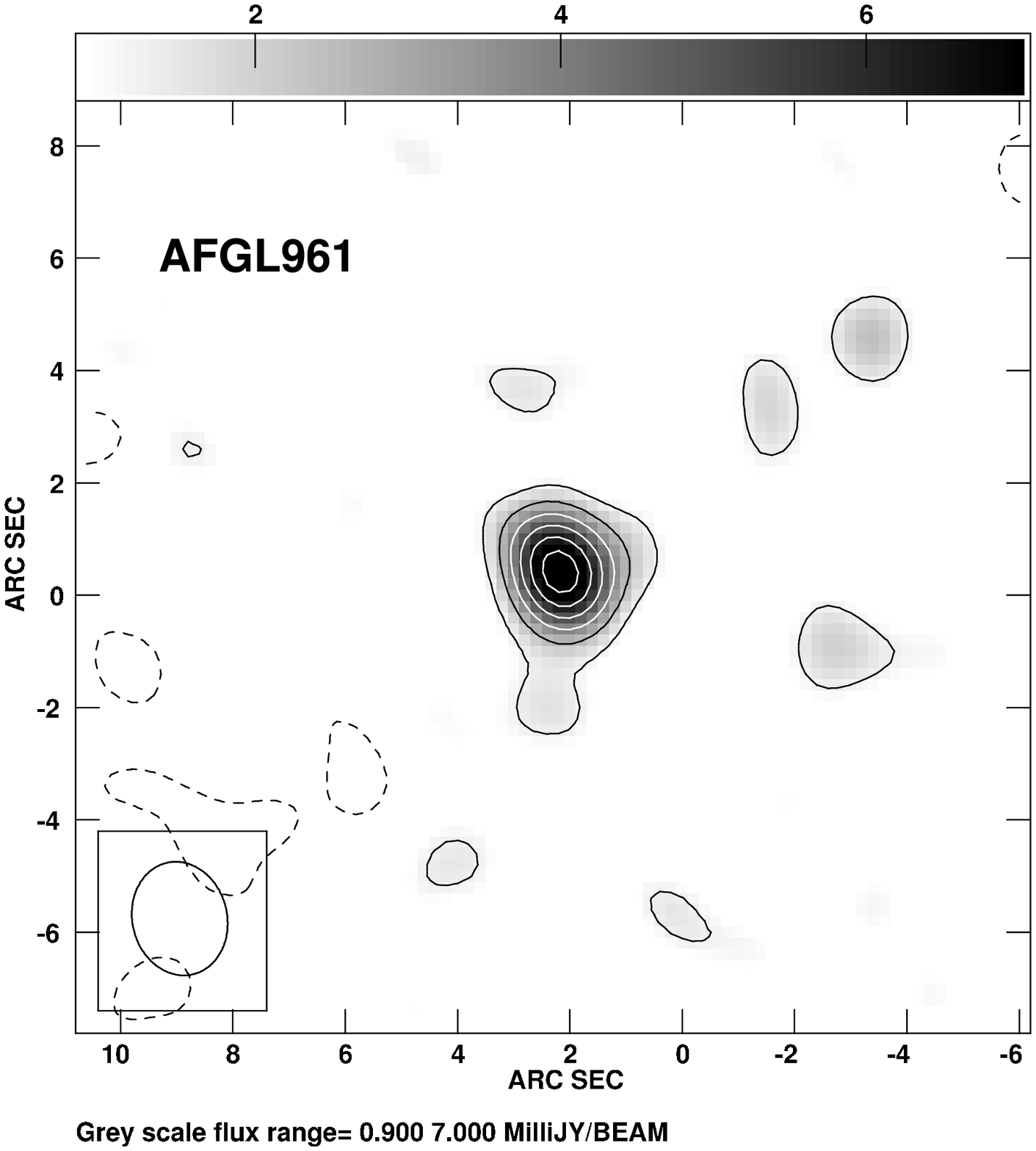}\\
\includegraphics[width=54mm,height=54mm]{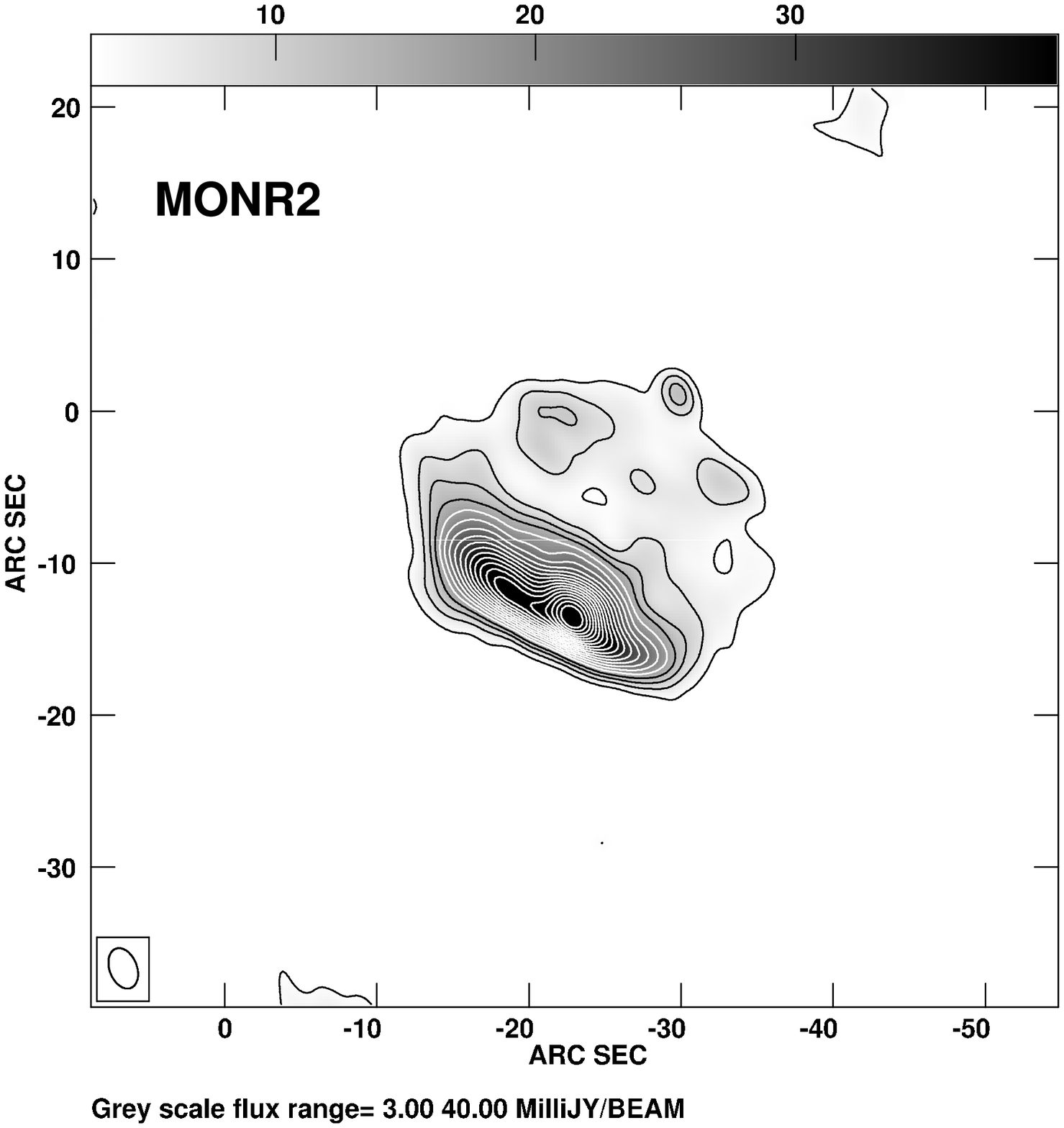}
\includegraphics[width=54mm,height=54mm]{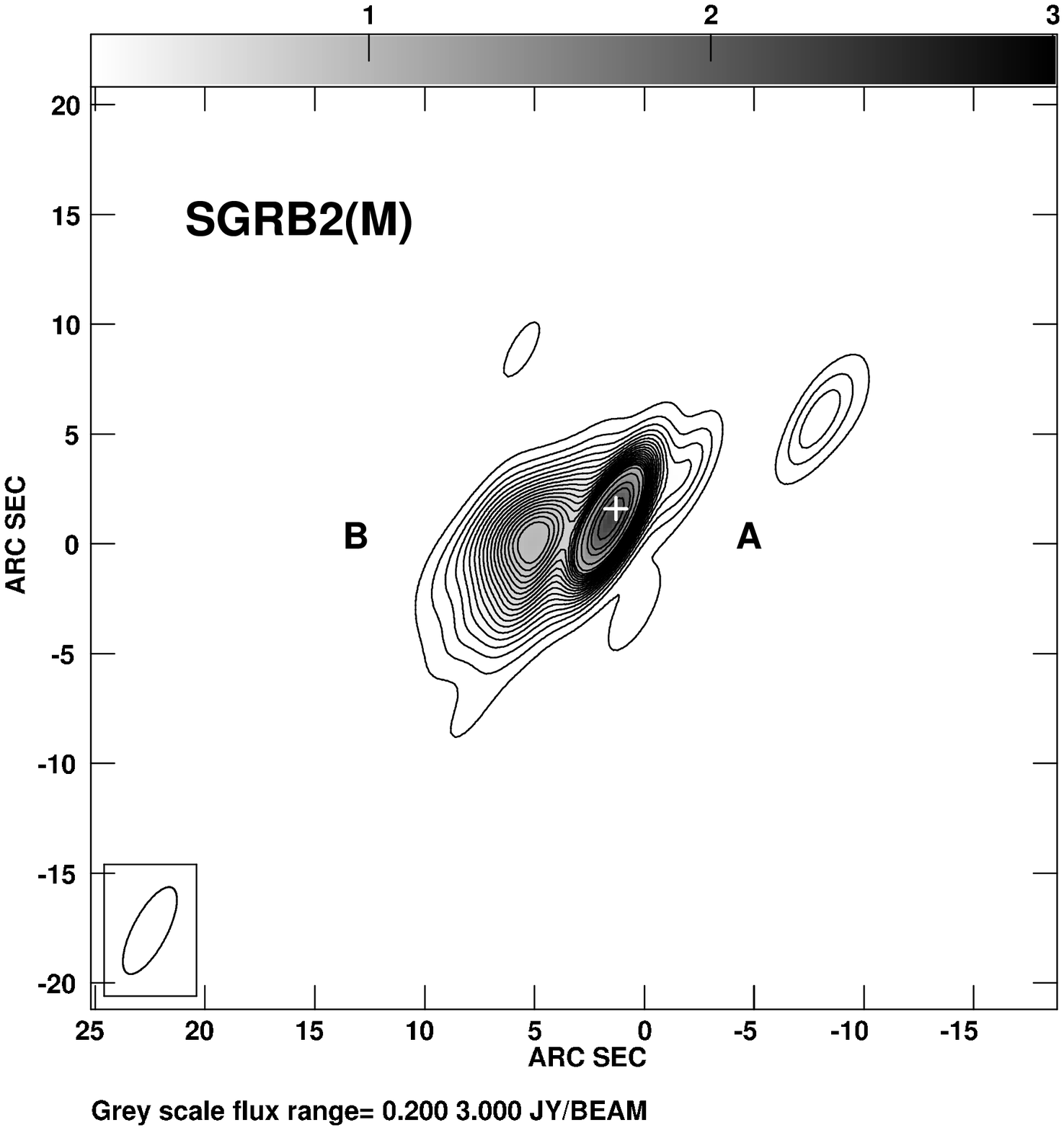}
\includegraphics[width=54mm,height=54mm]{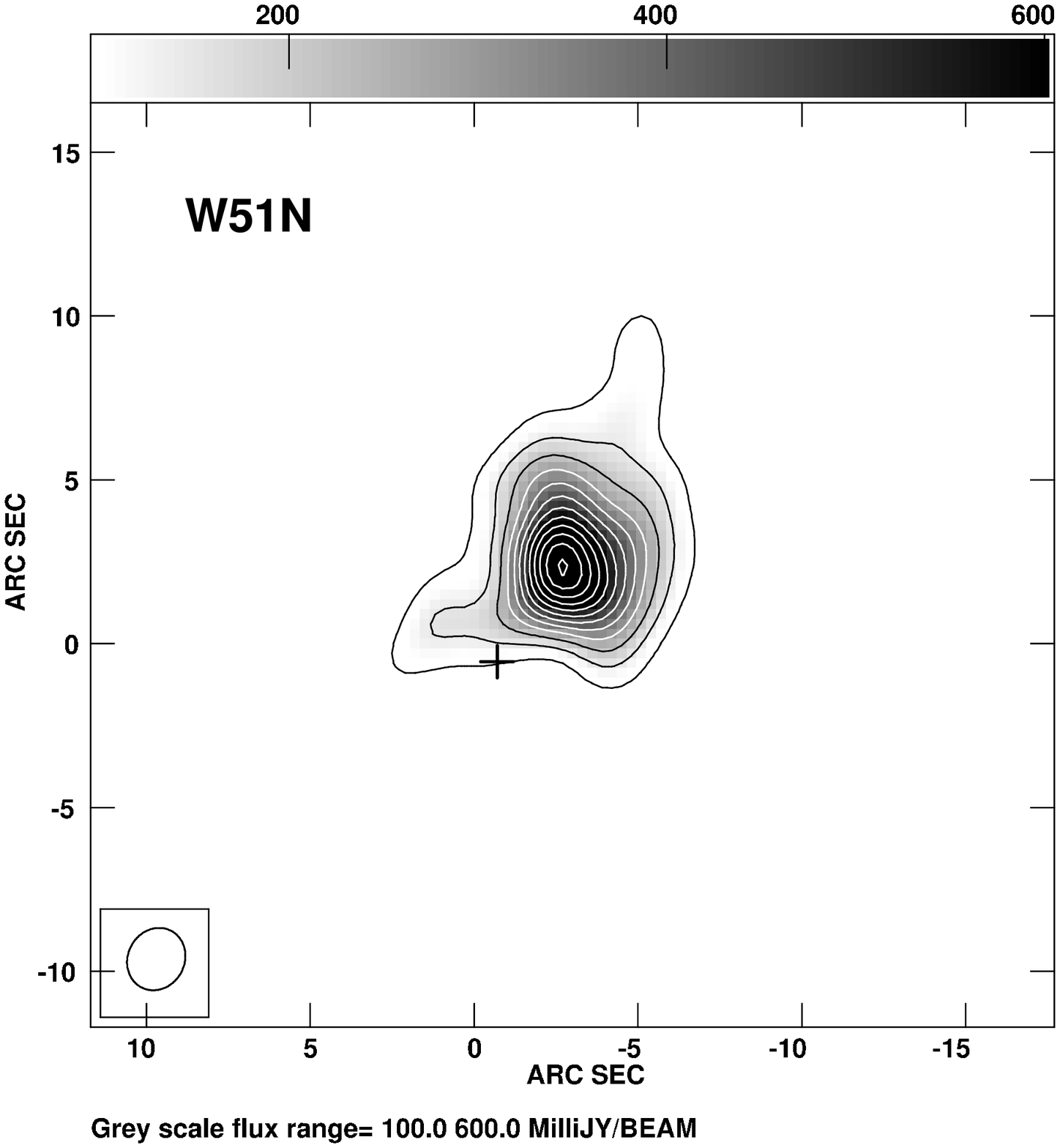}
\end{center}
\caption{Continuation.}
\end{figure}

\begin{figure}
\begin{center}
\includegraphics[angle=0,scale=.7]{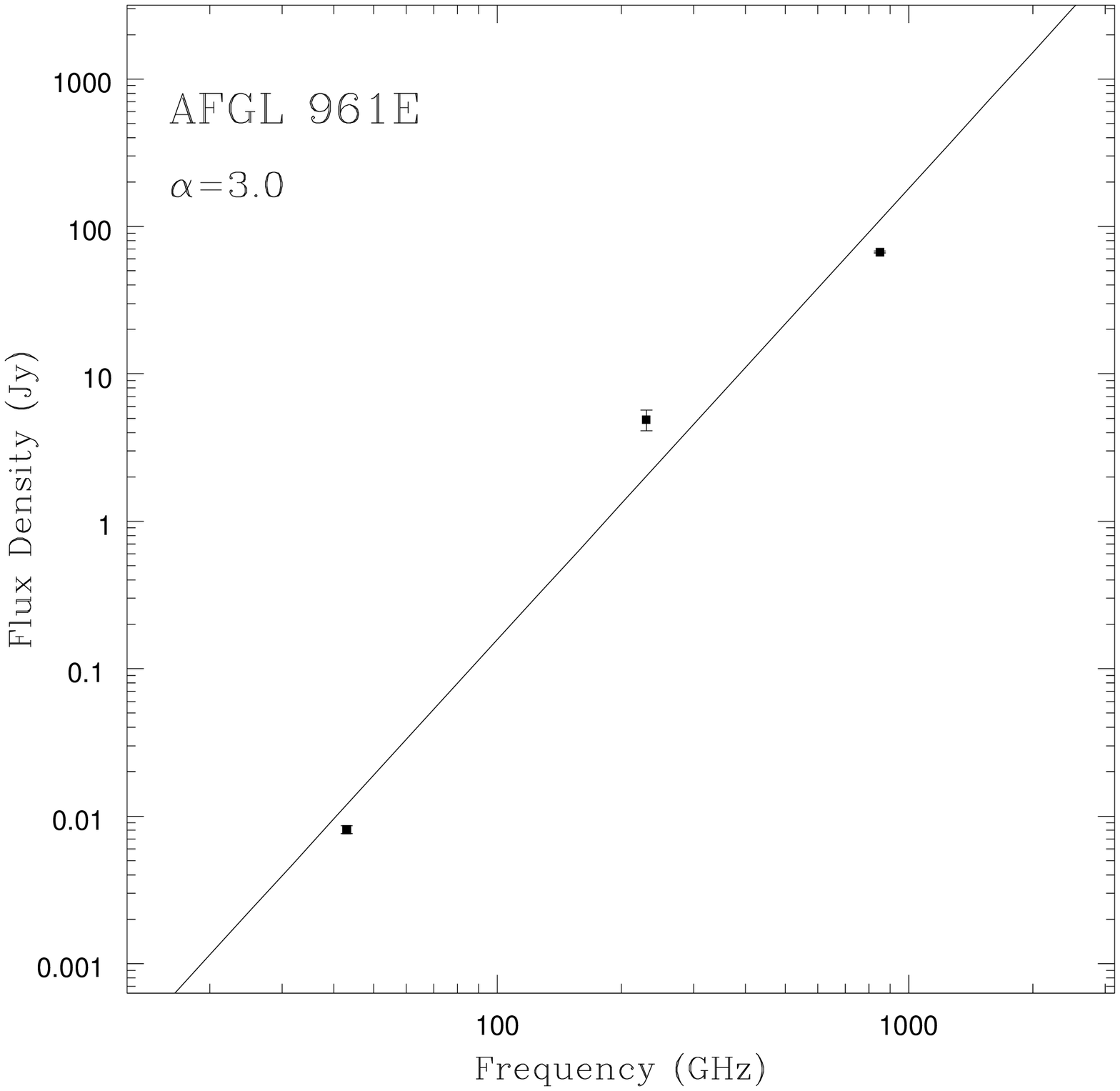}
\end{center}
\caption{Spectral energy distribution of the source AFGL 961E
 combining data obtained with the VLA (7 mm, this paper) and
 millimeter telescopes \citep{Guertleretal1991}).  The line is a
 least-squares power-law fit of the form (S$_\nu$ $\propto$
 $\nu^\alpha$) to the spectrum.\label{f4}}
\end{figure}

\begin{deluxetable}{l c c c c c r  c c}
\footnotesize
\tabletypesize{\scriptsize}
\tablecolumns{9} \tablewidth{0pc}
\tablecaption{\label{Params}Physical Parameters of Our Sample and
SiO and Continuum Emission Results}
\tablehead{
\colhead{} &
\multicolumn{2}{c}{Coordinates} &
\multicolumn{3}{c}{Physical Parameters} \\
\cline{4-6}
\colhead{} &
\colhead{$\alpha_{2000}$} &
\colhead{$\delta_{2000}$} &
\colhead{$D$} & 
\colhead{$L_{\rm bol}$} & 
\colhead{$M$} & 
\colhead{$\Delta S_{\rm M}$} & 
\colhead{$\Delta S_{\rm C}$} &
\colhead{} \\
\colhead{Source} &
\colhead{[$^h$  $^m$  $^s$]} &  
\colhead{[$^\circ$  $'$  $''$]} &
\colhead{(kpc)} &
\colhead{(L$_{\odot}$)}&
\colhead{(M$_{\odot}$)} &
\colhead{(mJy)} &
\colhead{(mJy)} &
\colhead{Ref.} }
\startdata
N6334I & 17 20 53.50 & -35 47 01.0 & 1.7 & 2.6 $\times$ 10$^5$ & 200 &
140 & \textbf{I}&1,2\\ N6334I-N & 17 20 54.60 & -35 45 08.0 & 1.7 &
1.9 $\times$ 10$^3$ & 400 & 140 & 10 &1,2\\ Sgr B2(M) & 17 47 20.00 &
-28 23 05.0 & 7.1 & 1 $\times$ 10$^6$ & 500 & \textbf{D}& \textbf{I} &
3,29\\ AFGL 2046 & 18 00 30.40 & -24 03 60.0 & 4.0 & 7 $\times$ 10$^5$
& 30 & 340 & \textbf{I}&4\\ I18426$-$0204 & 18 45 12.80 & -02 01 12.0
& 1.1 & 6.31 $\times$ 10$^2$ & 36 & 80& \textbf{I}& 5,6 \\ AFGL 7009S
& 18 34 19.70 & -05 59 44.0 & 3.0 & 2.9 $\times$ 10$^4$ &-- & 30&
\textbf{I}&25 \\ I18437$-$0216 & 18 46 22.70 & -02 13 24.0 &-- &-- &--
& 40& 10 &5,6 \\ I18440$-$0148 & 18 46 36.30 & -01 45 23.0 &-- &-- &--
& 40&\textbf{D}&5,6 \\ AFGL 2059 & 18 04 53.60 & -24 26 42.0 & 1.5 &
1.6 $\times$ 10$^4$ & -- & 40&3&14 \\ I18102$-$1800 & 18 13 12.20 &
-17 59 35.0 & 2.6 & 6.3 $\times$ 10$^4$ & 413 & 40&3&5,6 \\
G12.890+0.48 & 18 11 51.30 & -17 31 29.0 & 3.6 & 3.2 $\times$ 10$^4$ &
1200 & 50&3 &5,6\\ I18090$-$1832 & 18 12 01.90 & -18 31 56.0 & 6.6 &
1.2 $\times$ 10$^4$&1011 & 50&3& 5,6 \\ W33 A & 18 14 39.90 & -17 51
59.0 & 4.0 & 1 $\times$ 10$^5$ & 1089 & 50&3 & 11,13,14\\
I18223$-$1243 & 18 25 10.90 & -12 42 17.0 & 3.7 & 1.6 $\times$ 10$^4$
&527 & 40&3 & 5,6\\ I18151$-$1208 & 18 17 57.10 & -12 07 22.0 &-- &--
&-- & 80&3 &5,6\\ I18159$-$1550 & 18 18 47.30 & -15 48 58.0 & 4.7 &
1.5 $\times$ 10$^4$ & 393 & 80&3 &5,6\\ I18182$-$1433 & 18 21 07.90 &
-14 31 53.0 & 4.5 & 2 $\times$ 10$^4$ & 1507 & 60&3 &5,6\\ AFGL 2136 &
18 22 26.50 & -13 30 15.0 & 2.0 & 7 $\times$ 10$^4$ & - &
40&\textbf{I} &14\\ I18306$-$0835 & 18 33 21.80 & -08 33 38.0 & 4.9 &
1.3 $\times$ 10$^4$ & 1721 & 40&3 &5,6\\ W51N & 19 23 40.10 & 14 31
06.0 & 7 & 2.8 $\times$ 10$^6$ & 1000 & \textbf{D}
&\textbf{I}&15,16,17\\ W51E & 19 23 44.00 & 14 30 30.0 & 7 & -- & -- &
40 &\textbf{I}&15,16,17\\ I19266+1745 & 19 28 54.00 & 17 51 56.0 & 0.3
& 5 $\times$ 10$^1$ & 3 & 30&10&5,6\\ I19220+1432 & 19 24 19.70 & 14
38 03.0 &-- &-- &-- & 30&10&5,6\\ I19410+2258 & 19 43 11.40 & 23 44
06.0 &2.1 & 1 $\times$ 10$^4$ & 418 & 30&3&5,6\\ I19411+2306 & 19 43
18.10 & 23 13 59.0 &2.9 & 5 $\times$ 10$^3$ & 245 & 20&3&5,6 \\
I19471+2641 & 19 49 09.90 & 26 48 52.0 &-- &-- &-- & 40&4&5,6 \\ NGC
6820 & 19 42 27.90 & 23 05 15.0 &2.4 &6.3 $\times$ 10$^3$ &65 &
30&2&5,6 \\ I19413+2306 & 19 43 28.90 & 23 40 04.0 &1.8 & 2 $\times$
10$^2$ & 115 & 30&10&5,6 \\ I20051+3435 & 20 07 03.80 & 34 44 35.0
&1.6 & 2 $\times$ 10$^2$ & 48 & 60&4&5,6 \\ I20081+2720 & 20 10 11.50
& 27 29 06.0 &-- &-- &-- & 40&4&5,6 \\ I20126+4104 & 20 14 26.00 & 41
13 32.0 &-- &-- &-- & 20&3&5,6 \\ I20205+3948 & 20 22 21.90 & 39 58
05.0 &-- &-- &-- & 40&3&5,6 \\ CPM 35 & 20 23 23.80 & 41 17 40.0 &1.7
& 1.9 $\times$ 10$^2$ & 171 & 30&3&5,6 \\ AFGL 2591 & 20 29 24.90 & 40
11 21.0 & 1.0 & 2 $\times$ 10$^4$ & 42 & 30&\textbf{I}&14 \\
I20293+3952 & 20 31 10.70 & 40 03 10.0 &1.3 & 2 $\times$ 10$^3$ & 98 &
30&4&5,6 \\ I20319+3958 & 20 33 49.30 & 40 08 45.0 &-- &-- &-- &
30&4&5,6 \\ I20332+4124 & 20 35 00.50 & 41 34 48.0 &-- &-- &-- &
20&4&5,6 \\ I20343+4129 & 20 36 07.10 & 41 40 01.0 &-- &-- &-- &
30&8&5,6 \\ W75 N & 20 38 36.40 & 42 37 22.0 & 2.0 & a few $\times$
10$^5$ & -- & 30&\textbf{I}& 18\\ DR21(OH) & 20 39 00.90 & 42 22 38.0
& 3.0 & 1 $\times$ 10$^3$ & 350 & 20&\textbf{I}&12,13 \\ I22134+5834 &
22 15 09.10 & 58 49 09.0 &-- &-- &-- & 20&4&5,6 \\ AFGL 2884 & 22 19
18.10 & 63 18 49.0 & 0.9 &-- &-- & 20&\bf{I}&28 \\ I22551+6221 & 22 57
05.20 & 62 37 44.0 &-- &-- &-- & 20&4&5,6 \\ I22570+5912 & 22 59 06.50
& 59 28 49.0 &-- &-- &-- & 30&4&5,6 \\ I23033+5951 & 23 05 25.70 & 60
08 08.0 &-- &-- &-- & 30&4&5,6 \\ NGC 7538 IRS1 & 23 13 45.40 & 61 28
11.0 & 2.8 &1.3 $\times$ 10$^5$ & 815 & 30&\textbf{I}&9,10,13\\ NGC
7538 IRS9 & 23 14 01.70 & 61 27 20.0 & 2.8 &4.0 $\times$ 10$^4$ & 430
& 30&4&9,10,13\\ G111.25 & 23 16 09.30 & 59 55 23.0 & 4.8 &2.0
$\times$ 10$^4$ & -- & 30&4&27 \\ I23151+5912 & 23 17 21.00 & 59 28
28.0 &-- &-- &-- & 40&3&5,6\\ I23545+6508 & 23 57 05.20 & 65 25 11.0
&-- &-- &-- & 40&3&5,6\\ W3 IRS5 & 02 25 40.70 & 62 05 52.0 & 2.2 &
1.7 $\times$ 10$^5$ & 262 & 60&\textbf{I}&7,8,13 \\ W3(OH) & 02 27
04.50 & 61 52 25.0 & 2.0 & 1 $\times$ 10$^5$ & 50 & 30&\textbf{I}&21\\
AFGL 490 & 03 27 38.40 & 58 47 05.0 & 1.0 & 2.2-4.0 $\times$ 10$^3$&
490 & 30&\textbf{I}&20\\ S255 & 06 12 54.90 & 17 59 23.0 & 2.5 &6
$\times$ 10$^4$ & -- & 40&3&23\\ G192.16 & 05 58 13.90 & 16 31 60.0 &
2.0 & 1 $\times$ 10$^3$ & -- & 50&3&26\\ ORION-I & 05 35 14.50 & -05
22 30.0 &0.41 &$\sim$ 10$^5$ & -- &\textbf{D}&16&19\\ I05358+3543 & 05
39 10.40 & 35 45 19.0 &-- &-- &-- & 40&3&5,6\\ I05490+2658 & 05 52
12.90 & 26 59 33.0 &-- &-- &-- & 40&3&5,6\\ AFGL 989 & 06 41 10.00 &
09 29 19.0 &0.8 & 2 $\times$ 10$^3$ & -- & 50&3&22\\ AFGL 961 & 06 34
37.60 & 04 12 44.0 &1.6 & 8.9 $\times$ 10$^3$ & -- &
20&\textbf{I}&24\\ MonR2 & 06 07 47.80 & -06 22 55.0 &1.0 & 1.3
$\times$ 10$^4$ & -- & 30&\textbf{I}&14\\
\enddata
\tablecomments{ \footnotesize
The first column gives the source designation. I as first character in
a source name stands for IRAS.  Second and third column give right
ascension and declination (J2000), respectively. Forth column lists
the distance where the near kinematic distance is taken for sources
discussed in refs. 5 and 6.  Fifth and sixth columns the bolometric
luminosity of the region containing the source and $M$ the mass of its
surrounding molecular core, respectively.  Seventh column: $\Delta
S_{\rm M}$ is 4 times the rms noise in a single 2.72 \kms\ wide spectral channel. 
The noise levels of the $v=1$ and 2 data
cubes were very similar.
\textbf{D} indicates that maser mission was detected (see Table \ref{Masers} and Figure 1).
Eight column: $\Delta S_{\rm C}$ is 4 times the rms noise of the
continuum images.
\textbf{I} indicates that continuum emission was detected and imaged 
(for these sources, see Table \ref{Continuum} and Figure 2). 
References are listed in the ninth column. They are: 1:
\citet{Sandell2000}; 2: \citet{Neckel1978}; 3: \citet{Reidetal1988};
4: \citet{Gomezetal1991} and references therein; 5:
\citet{Beutheretal2002}; 6: \citet{Sridharanetal2002}; 7:
\cite{Laddetal1993}; 8: \cite{Humphreys1978}; 9:
\cite{Werneretal1979}; 10: \cite{Cramptonetal1978}; 11:
\cite{Guertleretal1991}; 12: \cite{vanderTaketal1999}; 13:
\cite{vanderTaketal2000}; 14: \cite{BoonmanvanDishoeck2003} and references therein; 15:
\cite{Schnepsetal1981}; 16: \cite{EricksonTokunaga1980}; 17:
\cite{Rudolphetal1990}; 18: \cite{Torrellesetal1997}; 19:
\cite{MentenReid1995}; 20: \cite{Schreyeretal2002}; 21:
\cite{Wrightetal2004} and references therein; 22: \cite{Perettoetal2006} 
and references therein; 23:
\cite{AlvarezHoare2005} and references therein; 24: \cite{Kleinetal2005}; 25:
\cite{Zavagnoetal2002}; 26: \cite{Shepherdetal2004}; 27:
\cite{Trinidadetal2006} and references therein; 28: \cite{Kurtzetal1994};
29: \cite{DeVicenteetal2000}\label{table1}}
\end{deluxetable}

\begin{deluxetable}{l c c c c c}
\tablecolumns{6}
\scriptsize
\tablewidth{0pc}
\tablecaption{Parameters of the Phase Calibrators}
\tablehead{
\colhead{}                       &
\multicolumn{2}{c}{Phase Center} &
\colhead{Bootstrapped}           &
\multicolumn{2}{c}{Synthesized Beam} \\
\cline{5-6}
\cline{2-3}
\colhead{}                       &
\colhead{$\alpha_{2000}$}  &
\colhead{$\delta_{2000}$}  &
\colhead{Flux Density} &
\colhead{Size}         &
\colhead{P.A.}                \\
\colhead{Calibrator} &
\colhead{[$^h$  $^m$  $^s$]}  &
\colhead{[$^\circ$  $'$  $''$]}  &
\colhead{[Jy]}  &
\colhead{[arcsec]}   &
\colhead{[deg.] }
}
\startdata
1717-337 & 17 17 36.03 &-33 42 08.7& 8.9 & 5.44 $\times$ 1.46 & -25.3
\\ 1733-130 & 17 33 02.70 &-13 04 49.5& 7.9 & 2.53 $\times$ 1.62 &
-22.5 \\ 1833-210 & 18 33 39.91 &-21 03 40.0& 16.2& 2.48 $\times$ 1.42
& -11.7 \\ 1851+005 & 18 51 46.72 & 00 35 32.4& 1.8 & 2.03 $\times$
1.60 & -30.0 \\ 1832-105 & 18 32 20.83 &-10 35 11.2& 2.0 & 1.99
$\times$ 1.44 & -1.60 \\ 1743-038 & 17 43 58.85 &-03 50 04.6& 6.6 &
2.13 $\times$ 1.47 & -23.8 \\ 1925+211 & 19 25 59.60 & 21 06 26.1& 2.1
& 1.47 $\times$ 1.34 & 0.0 \\ 2015+371 & 20 15 28.73 & 37 10 59.5& 3.8
& 1.51 $\times$ 1.42 & -45.3 \\ 22250+558 & 22 50 42.85 & 55 50 14.5&
1.1 & 1.63 $\times$ 1.43 & -29.3\\ 0228+673 & 02 28 50.05 & 67 21
03.0& 5.2 & 1.89 $\times$ 1.42 & -49.0 \\ 0613+131 & 06 13 57.69 & 13
06 45.4& 1.7 & 1.57 $\times$ 1.48 & -3.4 \\ 0541-056 & 05 41 38.08
&-05 41 49.4& 1.5 & 1.92 $\times$ 1.49 & 0.3 \\ 0555+398 & 05 55 30.80
& 39 48 49.1& 3.5 & 1.54 $\times$ 1.38 & -10.6 \\
\enddata
\label{caltable}
\tablecomments{\footnotesize Positional accuracy is estimated to be 0.01$''$}
\end{deluxetable}

\begin{deluxetable}{l l c c l c c c c}
\tablecolumns{7}
\tabletypesize{\scriptsize}
\tablewidth{0pc}
\tablecaption{\label{Masers}Properties of the Detected Maser Emission}
\tablehead{
\colhead{Source} &
\multicolumn{2}{c}{Peak Position} &
\colhead{Line} &
\colhead{$S_{\rm p}$}                       &
\colhead{\textit{v}$_{\rm p}$}  &
\colhead{$\int S dv$}                       &
\colhead{\textit{v-}range}                       &
\colhead{$L_\nu$}           \\
\cline{2-3}
&
\colhead{$\alpha_{2000}$}  &
\colhead{$\delta_{2000}$}  &
&
\colhead{(Jy~beam$^{-1}$)} &
\colhead{(\kms)}         &
\colhead{(Jy~km~s$^{-1}$)} &
\colhead{(\kms)}              &
\colhead{(s$^{-1}$)}\\
\colhead{} &
\colhead{[$^h$  $^m$  $^s$]} &  
\colhead{[$^\circ$  $'$  $''$]} &
\colhead{} &
\colhead{} &
\colhead{} &
\colhead{} &
\colhead{} 
}
\startdata
Source-I$^{a}$ &05 35 14.507&-05 22 30.47 & $v=1$ &490 &$-3$ &
6463(35) &$ [-13,23]$& $6.7\times10^{44}$\\ 
&05 35 14.505&-05 22 30.42 & $v=2$ &279 &$-8$ & 2712(26) &$ [-13,23]$& $2.8\times10^{44}$\\ 
Sgr B2(M) &17 47 20.098& -28 23 03.40 & $v=1$ &\dd0.93 & 91 &
\dd\d8.8(0.4) &$ [81,98]$& $3.4\times10^{44}$\\ 
&17 47 20.095& -28 23 03.53 & $v=2$ &\dd0.19 & 96 & \dd\d0.88(0.08)&$[\approx79,98]$& $3.4\times10^{43}$\\
W51N &19 23 40.055&14 31 05.51 & $v=1$&\dd1.0 & 49 & \dd\d5.1(0.1) &$ [45,56]$& $1.5\times10^{44}$\\ 
&19 23 40.052&14 31 05.45 & $v=2$ &\dd2.5 & 49 & \dd15(0.1) &$ [45,61]$& $4.4\times10^{44}$\\ 
Miras & & & $v=1$ & & & & & a few $\times10^{43}$\\ 
& & & $v=2$ & & & & & a few $\times10^{43}$\\ 
RSGs & & & $v=1$ & & & & & $10^{44}$--$10^{46}$\\ 
 & & & $v=2$ & & & & & $10^{44}$--$10^{46}$\\ 
\enddata
\tablecomments{\footnotesize Measured values are given for the SiO $J=1-0, v=1$ and $v=2$ lines.
$S_{\rm p}$ is the flux density at the velocity of peak emission,
\textit{v}$_{\rm p}$.  $\int S dv$ is flux density integrated over the
\textit{v-}range showing emission. $L_\nu$ is the isotropic luminosity
in units of photons per second. For the 43.3 GHz SiO $J=1-0$ lines
$L\nu = 10^{44}$ s$^{-1}$ corresponds to a luminosity of
$7.5\times10^{-6}$ \Lsun.}
\tablecomments{\footnotesize Positional accuracy is estimated to be 0.01$''$}
\tablenotetext{a}{\footnotesize The peak position corresponds to the blueshifted integrated 
emission and the redshifted peak position is $\alpha[J2000]$=05$^h$35$^m$14.500$^s$ and
$\delta[J2000]$=-05$^\circ$22$'$30.36$''$ for $v=1$ and $v=2$.}
\end{deluxetable}

\clearpage

\begin{deluxetable}{l c c c c c c c c}
\centering
\tabletypesize{\scriptsize}
\tablecolumns{9}
\tablewidth{0pc}
\tablecaption{\label{Continuum}Parameters of the 43.3 GHz Continuum Sources}
\tablehead{
\colhead{}                       &
\multicolumn{2}{c}{Peak Position} &
\colhead{Density}                &
\colhead{Peak} &
\multicolumn{2}{c}{Deconvolved Parameters} &\\
\cline{6-8}
\cline{2-3}
\colhead{}           &
\colhead{$\alpha_{2000}$}  &
\colhead{$\delta_{2000}$}  &
\colhead{Flux} &
\colhead{Flux} &
\colhead{Size}   &
\colhead{P.A.}   & \\
\colhead{Source} &
\colhead{[$^h$  $^m$  $^s$]}  &
\colhead{[$^\circ$  $'$  $''$]}  &
\colhead{[$m$Jy]}  &
\colhead{[$m$Jy Beam$^{-1}$]}  &
\colhead{[arcsec$^2$]}  &
\colhead{[deg.] }
}
\startdata
   N6334I & 17 20 53.428 & -35 47 01.60 & 2300 & 1400 & 5.7 $\times$
   2.6 & -32.0 \\ Sgr B2(M)-A & 17 47 20.121 & -28 23 03.90 & 3600 &
   2600 & 2.1 $\times$ 1.2 & 146 \\ Sgr B2(M)-B & 17 47 20.380 & -28
   23 05.00 & 3200 & 1800 & 2.7 $\times$ 2.5 & 144 \\ AFGL 2046 & 18
   00 30.442 & -24 04 01.21 & 7600 & 2700 & 3.0 $\times$ 2.7 & 84 \\
   I18440 & 18 46 36.132 & -01 45 17.61 & 5.5 & 6.4 & - & - \\ AFGL
   2136 & 18 22 26.409 & -13 30 12.25 & 7.4 & 5.7 & - & - \\ W51N & 19
   23 39.917 & 14 31 09.19 & 3800 & 624 & 3.9 $\times$ 2.9 & 20\\
   W51e2 & 19 23 43.918 & 14 30 34.70 & 391 & 390 & - & - \\ W51e1 &
   19 23 43.786 & 14 30 26.32 & 42 & 39 & - & - \\ W51IRS1 & 19 23
   41.913 & 14 30 36.23 & 747 & 182 & 3.4 $\times$ 1.9 & 157 \\ AFGL
   2591 & 20 29 24.545 & 40 11 15.06 & 50 & 40 & 1.9 $\times$ 1.1 & 79
   \\ W75N-VLA3Bb & 20 38 36.475 & 42 37 33.93 & 24 & 18 & 2.1
   $\times$ 1.1 &174\\ DR21(OH) & 20 39 01.027 & 42 22 49.23 & 16 &
   6.9 & 3.0 $\times$ 1.3 & 2.8\\ AFGL 2884 & 22 19 18.169 & 63 18
   46.31 & 20 & 15 & 2.1 $\times$ 0.5 &38\\ NGC 7538 IRS1-A & 23 13
   45.381& 61 28 10.18 & 530 & 530 & - & - \\ NGC 7538 IRS1-B & 61 28
   18.806& 23 13 45.39 & 710 & 82 & 4.5 $\times$ 4.2 & 153\\ W3 IRS5 &
   02 25 37.900 & 62 05 50.48 & 520 & 160 & - & - \\ W3(OH) & 02 27
   03.887 & 61 52 24.54 & 1500 & 866 & - & - \\ AFGL 490 & 03 27
   38.834 & 58 46 59.95 & 9.8 & 8.2 & - & - \\ AFGL 961 & 06 34 37.743
   & 04 12 44.43 & 8.1 & 7.9 & - & - \\ MonR2 & 06 07 46.270 & -06 23
   08.55 & 940 & 730 & - & -
\enddata
\tablecomments{\footnotesize Those parameters were determined from a linearized
                             least-squares fit to a Gaussian ellipsoid
                             function using the AIPS task IMFIT. Sources with 
                             failed deconvolved fitting are indicated with the sign '-'.}
\tablecomments{\footnotesize Positional accuracy is estimated to be 0.01$''$}

\end{deluxetable}

\clearpage

\begin{deluxetable}{l c c c}
\centering
\tabletypesize{\scriptsize}
\tablecolumns{4}
\tablewidth{0pc}
\tablecaption{Nature of the 42 GHz Continuum Emission}
\tablehead{
\colhead{Name}                 &
\colhead{Nature}               &
\colhead{References}           &
}
\startdata
   N6334I & UC HII region (NGC 6334F) & \citet{Carraletal2002} \\ Sgr
   B2(M) & Cluster of HII regions &
   \citet{Gaumeetal1995};\citet{DePreeetal1996}\\ & &
   \citet{DePreeetal1998} \\ AFGL 2046 & UC HII region (G5.89-0.39) &
   \citet{Gomezetal1991} \\ IRAS 18440 & UC HII region? & Here \\ AFGL
   2136 & UC HII region? (RS4) & \citet{MentenvanderTak2004}\\ W51N &
   UC HII region(W51-IRS2/d2/d1) &
   \citet{Gaumeetal1993};\citet{ZhangHo1997}\\ W51-e2 & UC HII region
   & \citet{Gaumeetal1993};\citet{ZhangHo1997} \\ W51-e1 & UC HII
   region & \citet{Gaumeetal1993};\citet{ZhangHo1997} \\ W51-IRS1 &
   Compact HII region & \citet{Gaumeetal1993};\citet{ZhangHo1997}\\
   AFGL 2591 & UC HII region (VLA1)/Ionized thermal jet &
   \citet{Trinidadetal2003}; \citet{vanderTaketal2005} \\ W75 North &
   Ionized thermal jet (VLA3-Bb) & \citet{Shepherdetal2003}\\
   DR21(OH) & Dusty Core (MM1) & Here \\ AFGL 2884 & Ionized stellar
   wind?(S140 IRS 1) & \citet{Hoare2002};\citet{Hoare2006} \\
   N7538-IRS1-A & UCHII Variable & \citet{francoetal2003};
   Here; \\ & & \citet{WynnWilliamsetal1974} \\ N7538-IRS1-B & Compact
   HII region & \citet{WynnWilliamsetal1974}; Here \\ W3 IRS5 & UC HII
   region & \citet{MentenvanderTak2004} \\ W3(OH) & UC HII region &
   \citet{Wilneretal1999} \\ AFGL 490 & Circumstellar Disk/Envelope &
   \citet{Chinietal1991};\citet{Henningetal1990};
   \citet{Schreyeretal2006} \\ AFGL 961E & Dusty Core? & Here \\ MONR2
   & Compact HII region & Here \enddata
\end{deluxetable}
\end{document}